\documentclass{aa}
\usepackage{graphicx}
\usepackage{natbib}
\usepackage{color}
\usepackage{bm,url,times}
\graphicspath{{./fig/}{./png/}}

\newcommand{\EQ}{\begin{equation}}
\newcommand{\EN}{\end{equation}}
\newcommand{\EQA}{\begin{eqnarray}}
\newcommand{\ENA}{\end{eqnarray}}

\newcommand{\Eq}[1]{Eq.~(\ref{#1})}
\newcommand{\Eqs}[2]{Eqs.~(\ref{#1}) and~(\ref{#2})}
\newcommand{\Eqss}[2]{Eqs.~(\ref{#1})--(\ref{#2})}

\newcommand{\App}[1]{Appendix~\ref{#1}}
\newcommand{\Sec}[1]{Sect.~\ref{#1}}

\newcommand{\Fig}[1]{Fig.~\ref{#1}}

\newcommand{\Tab}[1]{Table~\ref{#1}}

{}
{}
{}

{}
{}
{}
{}
{}
{}
{}
{}
{}
{}
{}
{}
{}
{}
{}
{}
{}
{}
{}
{}
{}

{}
{}
{}

{}

{}
{}

\newcommand{\tkapz}{\tilde{\kappa_0}}

\newcommand{\PCn}{{\sc Pencil Code}}
\newcommand{\nnn}{\hat{\mbox{\boldmath $n$}} {}}

\newcommand{\kk}{\bm{k}}

\newcommand{\xx}{\bm{x}}

\newcommand{\uu}{\mbox{\boldmath $u$} {}}

\newcommand{\FF}{\mbox{\boldmath $F$} {}}

\newcommand{\grav}{\mbox{\boldmath $g$} {}}
\newcommand{\nab}{\mbox{\boldmath $\nabla$} {}}
\newcommand{\nabad}{\nabla_{\rm ad}}

\newcommand{\IIII}{\mbox{\boldmath ${\sf I}$} {}}

\newcommand{\SSSS}{\mbox{\boldmath ${\sf S}$} {}}

\newcommand{\DD}{{\rm D} {}}

\newcommand{\dd}{{\rm d} {}}

\newcommand{\const}{{\rm const}  {}}

\def\la{\mathrel{\mathchoice {\vcenter{\offinterlineskip\halign{\hfil
$\displaystyle##$\hfil\cr<\cr\sim\cr}}}
{\vcenter{\offinterlineskip\halign{\hfil$\textstyle##$\hfil\cr<\cr\sim\cr}}}
{\vcenter{\offinterlineskip\halign{\hfil$\scriptstyle##$\hfil\cr<\cr\sim\cr}}}
{\vcenter{\offinterlineskip\halign{\hfil$\scriptscriptstyle##$\hfil\cr<\cr\sim\cr}}}}}
\def\ga{\mathrel{\mathchoice {\vcenter{\offinterlineskip\halign{\hfil
$\displaystyle##$\hfil\cr>\cr\sim\cr}}}
{\vcenter{\offinterlineskip\halign{\hfil$\textstyle##$\hfil\cr>\cr\sim\cr}}}
{\vcenter{\offinterlineskip\halign{\hfil$\scriptstyle##$\hfil\cr>\cr\sim\cr}}}
{\vcenter{\offinterlineskip\halign{\hfil$\scriptscriptstyle##$\hfil\cr>\cr\sim\cr}}}}}

\newcommand{\sigmaSB}{\sigma_{\rm SB} }

\def\Ra{\mbox{\rm Ra}}
\def\Ma{\mbox{\rm Ma}}

\def\Pra{\mbox{\rm Pr}}

\def\Pe{\mbox{\rm Pe}}

\def\Pet{\widetilde {\rm Pe}}

\def\csz{c_{\rm s0}}
\def\cp{c_{\rm p}}
\def\cv{c_{\rm v}}

\def\Teff{T_{\rm eff}}
\def\cs{c_{\rm s}}

\def\Hp{H_{\rm p}}

\def\urms{u_{\rm rms}}

\def\half{{\textstyle{1\over2}}}

\def\onethird{{\textstyle{1\over3}}}

\newcommand{\K}{\,{\rm K}}
\newcommand{\g}{\,{\rm g}}
\newcommand{\s}{\,{\rm s}}

\newcommand{\ks}{\,{\rm ks}}
\newcommand{\cm}{\,{\rm cm}}

\newcommand{\km}{\,{\rm km}}

\newcommand{\Mm}{\,{\rm Mm}}

\newcommand{\erg}{\,{\rm erg}}
\newcommand{\mol}{\,{\rm mol}}

\newcommand{\yapj}[3]{ #1, {ApJ,} {#2}, #3}

\newcommand{\yapjl}[3]{ #1, {ApJL,} {#2}, #3}

\newcommand{\yana}[3]{ #1, {A\&A,} {#2}, #3}

\newcommand{\ypasj}[3]{ #1, {Publ.\ Astron.\ Soc.\ Japan,} {#2}, #3}
\newcommand{\yjfm}[3]{ #1, {J.\ Fluid Mech.,} {#2}, #3}

\newcommand{\ymn}[3]{ #1, {MNRAS,} {#2}, #3}

\newcommand{\ysph}[3]{ #1, {Solar Phys.,} {#2}, #3}

\newcommand{\yjcp}[3]{ #1, {J.\ Comput.\ Phys.,} {#2}, #3}
\newcommand{\yjour}[4]{ #1, {#2}, {#3}, #4}

\newcommand{\ybook}[3]{ #1, {#2} (#3)}
\newcommand{\yproc}[5]{ #1, in {#3}, ed.\ #4 (#5), #2}

\newcommand{\smn}[2]{ #1, {MNRAS}, submitted, arXiv:#2}
\title{Near-polytropic stellar simulations with a radiative surface}
\author{A. Barekat\inst{1,2,3} \and A. Brandenburg\inst{1,2}}
\date{Submitted August 8, 2013; Accepted May 30, 2014}
\institute{
Nordita, KTH Royal Institute of Technology and Stockholm University,
Roslagstullsbacken 23, 10691 Stockholm, Sweden
\and
Department of Astronomy, AlbaNova University Center,
Stockholm University, 10691 Stockholm, Sweden
\and
Max-Planck-Institut f\"ur Sonnensystemforschung, Justus-von-Liebig-Weg
3, 37077 G\"ottingen, Germany
}
\begin{document}

\abstract
{
Studies of solar and stellar convection often employ simple polytropic
setups using the diffusion approximation instead of solving the proper
radiative transfer equation.
This allows one to control separately the polytropic index of the
hydrostatic reference solution, the temperature contrast
between top and bottom, and the
Rayleigh and P\'eclet numbers.
}
{
Here we extend such studies by including radiative transfer in the
gray approximation using a Kramers-like opacity with freely adjustable
coefficients.
We study the properties of such models and compare them with results from
the diffusion approximation.
}
{
We use the {\sc Pencil Code}, which is a high-order finite difference
code where radiation is treated using the method of long characteristics.
The source function is given by the Planck function.
The opacity is written as $\kappa=\kappa_0\rho^a T^b$,
where $a=1$ in most cases, $b$ is varied from $-3.5$ to $+5$,
and $\kappa_0$ is varied by four orders of magnitude.
We adopt a perfect monatomic gas.
We consider sets of one-dimensional models and perform a
comparison with the diffusion approximation in one- and
two-dimensional models.
}
{
Except for the case where $b=5$, we find one-dimensional hydrostatic equilibria
with a nearly polytropic stratification and a polytropic index close to
$n=(3-b)/(1+a)$, covering both convectively stable ($n>3/2$) and unstable
($n<3/2$) cases.
For $b=3$ and $a=-1$, the value of $n$ is undefined a priori and the
actual value of $n$ depends then on the depth of the domain.
For large values of $\kappa_0$, the thermal adjustment time becomes
long, the P\'eclet and Rayleigh numbers become large, and the temperature
contrast increases and is thus no longer an independent input parameter,
unless the Stefan--Boltzmann constant is considered adjustable.
}
{
Proper radiative transfer with Kramers-like opacities provides a useful
tool for studying stratified layers with a radiative surface in ways that
are more physical than what is possible with polytropic models using
the diffusion approximation.
}
\maketitle

\section{Introduction}

Convection in stars and accretion disks is a consequence of
radiative cooling at the surface.
Pioneering work by \cite{Nor82,Nor85} has shown that realistic simulations
of solar granulation can be performed with not too much extra effort and
the required computing resources are comparable to the mandatory
costs for solving the hydrodynamics part.
Yet, many studies of hydrodynamic and hydromagnetic convection
today ignore the effects of proper radiative transfer, sometimes
even at the expense of using compute-intensive implicit solvers
to cope with a computationally stiff problem in the upper layers
where the radiative conductivity becomes large \citep[e.g.,][]{Cat91,GD08}.
Therefore, the main reason for ignoring radiation cannot be just
the extra effort, but it is more likely
a reduced flexibility in that one is confined
to a single physical realization of a system and the difficulty
in varying parameters that are in principle fixed by the physics.
With only a few exceptions \citep[e.g.,][]{Edw90}, radiation
hydrodynamics simulations of stratified convection also employ
realistic opacities combined with a realistic equation of state.
In the case of the Sun this means that one can only simulate
for the duration of a few days solar time \citep{SN89,SN98,SN12}.

There are other types of realistic simulations
that are able to cover longer time scales by simulating only
deeper layers, so they ignore radiation.
However, these simulations still need to pose an upper boundary
condition, where the gas is cooled \citep{Mie00}.
This leads to a granulation-like pattern at a depth
where the flow topology is known to consist of individual downdrafts
rather than a connected network of intergranular lanes.
This compromises the realism of such simulations.
Other types of simulations give up the ambition for realism altogether
and try to model a ``toy Sun'' in which the broad range of time and
length scales is compressed to a much narrower range \citep{KMCWB13}.
This can be useful if one wants to understand the physics of the solar
dynamo, where we are not even sure about the possible importance of
the surface \citep{Bra05}, or the physics of sunspots, where so far only
models of a toy Sun have produced spontaneous magnetic flux concentrations
similar to those of sunspots \citep{BKR13}.
It is therefore important to know how to manipulate the parameters
to accommodate the relevant physics, given certain numerical constraints
such as the number of mesh points available.

In the present paper we include radiation, which introduces the
Stefan--Boltzmann constant, $\sigmaSB$, as a new characteristic quantity
into the problem.
It characterizes the strength of surface cooling, or, conversely,
the temperature needed to radiate the flux that is transported
through the rest of the domain.
Earlier simulations that ignored radiation have specified the
surface temperature in an ad hoc manner so as to achieve a
certain temperature contrast across the domain.
An example are the simulations of \cite{Bra96}, who specified
a parameter $\xi$ as the ratio of pressure scale height at the surface,
which is proportional to the temperature at the top, and the
thickness of the convectively unstable layer.
Alternatively, one can use a radiative surface boundary condition.
It involves $\sigmaSB$ and couples therefore the surface temperature
$T_{\rm top}$ to the lower part of the system, so $T_{\rm top}$
is then no longer a free parameter, unless one chooses an effective
value of $\sigmaSB$ so as to achieve the desired temperature contrast. 
This was done in recent simulations by \cite{KMB12}, who kept
the aforementioned parameter $\xi$
as the basic control parameter, which then determines the effective
value of $\sigmaSB$ in their simulations.

The goal of the present work is to explore the physics of models
that introduce radiation without being confined to just one realization.
We do this by using a Kramers-like opacity law, but with freely
adjustable parameters.
It turns out that it is possible in some cases to imitate polytropic models
with any desired polytropic index and Rayleigh number.
This then eliminates any restrictions to a single setup,
allowing one to perform parameter surveys, just like with earlier
polytropic models.
To compare radiative transfer models with those
in the diffusion approximation, we consider
two-dimensional convection simulations.
An ultimate application of this work is to study the formation
of surface magnetic flux concentrations through the
negative effective magnetic pressure instability \citep{BKR13}
which has been able to produce already bipolar region \citep{JW13,MBKR14}
and to investigate the relation to the magnetic cooling instability
of \cite{KM00} that could favor sunspot formation in the presence
of radiative cooling.
This will be discussed again at the end of the paper.

We would like to point out that, in view of more general applications,
we cannot assume the effective temperature to be given or fixed.
Thus, unlike the case usually considered in the theory of stellar atmospheres,
the dependence of temperature on optical depth is not known a priori.
Therefore, it is more convenient to fix instead the temperature at
the bottom of the domain and obtain the effective temperature, and thus
the flux, as a result of the calculation.

We begin by presenting first the governing equations and then describe
the basic setup of our model.
Next we compare a set of one-dimensional simulations with the
associated polytropic
indices that correspond to Schwarzschild stable or unstable solutions.
Finally, we explore the effect of including radiative transfer
instead of using the diffusion approximation combined with
a radiative boundary condition by comparing one- and two-dimensional
simulations.

\section{The model}
\label{RT}

\subsection{Governing equations}

We solve the hydrodynamics equations for logarithmic density $\ln\rho$,
velocity $\uu$, and specific entropy $s$, in the form
\EQA
{\DD \ln \rho \over \DD t}&=&-\nab\cdot\uu, \\
 \rho{\DD \uu\over \DD t}&=&-\nab p +\rho\grav + \nab\cdot(2\rho\nu\SSSS), \\
 \rho T {\DD s \over \DD
  t}&=&-\nab\cdot\FF_{\rm rad}+2\rho\nu \SSSS^2,
\label{sRT} 
\ENA
where $p$ is the gas pressure, $\grav$ is the gravitational acceleration,
$\nu$ is the viscosity, 
$\SSSS=\half[\nab\uu+(\nab\uu)^T]-\onethird\IIII\nab\cdot\uu$
is the traceless rate-of-strain tensor, $\IIII$ is the unit tensor,
$T$ is the temperature,
and $\FF_{\rm rad}$ is the radiative flux.
For the equation of state, we assume a perfect gas with
$p=({\cal R}/\mu)T\rho$, where ${\cal R}$ is the universal gas constant
and $\mu$ is the mean molecular weight.
The pressure is related to $s$ via $p=\rho^\gamma\exp(s/\cv)$, where
the adiabatic index 
$\gamma=\cp/\cv$ is the ratio of specific heats at constant pressure
and constant volume, respectively, and $\cp-\cv={\cal R}/\mu$.
To obtain the radiative flux, we adopt the gray approximation,
ignore scattering, and assume that the source function $S$ (not
to be confused with the rate-of-strain tensor $\SSSS$) is given
by the frequency-integrated Planck function, so $S=(\sigmaSB/\pi)T^4$,
where $\sigmaSB$ is the Stefan--Boltzmann constant.
The divergence of the radiative flux is then given by 
\EQ
\nab\cdot\FF_{\rm rad}=-\kappa\rho \oint_{4\pi}(I-S)\,\dd\Omega,
\label{fff}
\EN
where $\kappa$ is the opacity per unit mass
(assumed independent of frequency) and $I(\xx,t,\nnn)$
is the frequency-integrated specific intensity corresponding to
the energy that is carried by radiation per unit area, per unit time,
in the direction $\nnn$, through a solid angle $\dd\Omega$.
We obtain $I(\xx,t,\nnn)$ by solving the radiative transfer equation,
\EQ
\nnn\cdot\nab I=-\kappa\rho\, (I-S),
\label{RT-eq}
\EN
along a set of rays in different directions $\nnn$ using the
method of long characteristics.

\subsection{Opacity}

For our work it is essential that we can control the value and
functional form of the opacity.
We therefore choose a Kramers-like opacity given by
\EQ
\kappa=\kappa_0\rho^a T^b,
\label{kappa}
\EN 
where $a$ and $b$ are free parameters that characterize
the relevant radiative processes.
It is useful to consider the radiative conductivity $K(\rho,T)$,
which is given by
\EQ
K(\rho,T)={16\sigmaSB T^3\over 3\kappa\rho}
={16\sigmaSB T^{3-b}\over 3\kappa_0\rho^{a+1}}.
\label{K-model}
\EN
We note that, in a plane-parallel polytropic atmosphere,
$T(z)$ varies linearly with height $z$ and in the stationary state,
$K(\rho,T)$ is constant in the optically thick part.
This implies that $\rho$ is proportional to $T^n$, where
\EQ
n={3-b\over 1+a}
\label{n_from_ab}
\EN
is the polytropic index
(not to be confused with the direction of the ray $\nnn$).
This relation was also used by \cite{Edw90}, but the author regarded
those solutions as `a little contrived'.
This is perhaps the case if such solutions are applied throughout the
entire domain.
It should also be noted that \cite{Edw90} included thermal conduction
along with radiative transfer.
This meant that one had to pose a boundary condition
for the temperature at the top also,
which will not be necessary in our case, where, unless stated otherwise,
no thermal conductivity is included.
Indeed, as we shall show, with a Kramers-like opacity,
nearly polytropic solutions are a natural outcome
in the lower optically thick part of the domain, while in the upper
optically thin part of the domain the stratification tends to become
approximately isothermal.

For a perfect gas, the specific entropy gradient is related to
the gradients of the other thermodynamic variables via
\EQ
\nab s=\cv\nab\ln p-\cp\nab\ln\rho=(n+1-\gamma n)\cv\nab\ln T,
\EN
and vanishes when $n=1/(\gamma-1)$.
For a monatomic gas where $\gamma=5/3$, the stratification is
Schwarzschild-stable for $n>3/2$.

\subsection{Boundary conditions}

We consider a slab with boundary conditions in the $z$ direction
at $z_{\rm bot}$ and $z_{\rm top}$, where we assume the
gas to be stress-free, i.e.,
\EQ
\partial u_x/\partial z=\partial u_y/\partial z=u_z=0
\quad\mbox{on $z=z_{\rm bot}$, $z_{\rm top}$}.
\EN
We assume zero incoming intensity at the top, and compute the incoming
intensity at the bottom from a quadratic Taylor expansion of the source
function, which implies that the diffusion approximation is obeyed;
see Appendix~A of \cite{HDNB06} for details.
To ensure steady conditions, we fix temperature at the bottom,
\EQ
T=T_{\rm bot}\quad\mbox{on $z=z_{\rm bot}$},
\EN
while the temperature at the top is allowed to evolve freely.
There is no boundary condition on the density, but since no mass
is flowing in or out, the volume-averaged density is
automatically constant.
Since most of the mass resides near the bottom, the density there
will not change drastically and will be close to its initial value
at the bottom.

\subsection{The radiation module}

We use for all simulations the
\PCn\footnote{{\url{http://pencil-code.googlecode.com/}}}, which solves
the hydrodynamic differential equations with a high-order
finite-difference scheme.
The radiation module was implemented by \cite{HDNB06}.
It solves the transfer equation in the form
\EQ
\dd I/\dd\tau=I-S,
\label{dI}
\EN
where $\dd\tau=\kappa\rho\,\dd l$ is the differential of the
optical depth along a given ray and $l$ is a coordinate along this ray.

The code is parallelized by splitting the calculation into parts that
are local and non-local with respect to each processor. 
There are two local
parts that are compute-intensive and one that is non-local and
fast, so it does not require any computation.
Since $S$ is assumed independent of $I$ (scattering is ignored),
we can write the solution of \Eq{dI} as an integral for $I(\tau)$,
which is thus split into two parts,
\EQ
I(\tau)=\underbrace{\int_0^{\tau_0}e^{\tau'-\tau}S(\tau')d\tau'}_{I_{\rm extr}}
+\underbrace{\int_{\tau_0}^{\tau}e^{\tau'-\tau}S(\tau')d\tau'}_{I_{\rm intr}},
\label{Itau}
\EN
where the subscripts `extr' and `intr' indicate respectively an extrinsic,
non-local contribution and an intrinsic, local one.
An analogous calculation is done for calculating $\tau$ along the
geometric coordinate as $\tau(l)=\int_0^{l_0}\kappa\rho\,\dd l'
+\int_{l_0}^l\kappa\rho\,\dd l'$, where $l_0$ is the geometric
end point on the previous processor.
In the first step, we calculate $I_{\rm intr}(\tau)$, which
can be evaluated immediately on all processors in parallel,
while the first integral is written in the form
$I_{\rm extr}(\tau)=I_0\,e^{\tau_0-\tau}$,
where $I_0$ and $\tau_0$ are already being computed as part
of the $I_{\rm intr}$ calculation on neighboring processors
and the results included in the last step of the computation.

In the second step, the values of
$\tau_0=\int_0^{l_0}\kappa\rho\,\dd l'$ and $I_0=I(\tau_0)$
are communicated from the end point of each ray on the previous processor,
which cannot be done in parallel, but this does not
require any computational time.
In the final step one computes
\EQ
I_{\rm extr}(\tau)=I_0 e^{\tau_0-\tau}
\EN
and constructs the final intensity as
$I(\tau)=I_{\rm extr}(\tau)+I_{\rm intr}(\tau)$.

Instead of solving the radiative transfer equation directly for the
intensity, the contribution to the cooling term $Q(\tau)=I(\tau)-S(\tau)$
is calculated instead, as was done also by \cite{Nor82}.
This avoids round-off errors in the optically thick part.
For further details regarding the implementation we refer to \cite{HDNB06}.
To avoid interpolation, the rays are chosen such that they
go through mesh points.
The angular integration in \Eq{fff} is discretized as
\EQ
\nab\cdot\FF_{\rm rad}=-{4\pi\kappa\rho\over N}\,{D\over3}
\sum_{i=1}^N[I(\xx,t,\nnn_i)-S],
\label{Sum}
\EN
where $i$ enumerates the $N$ rays with directions $\nnn_i$ and $D/3$
is a correction factor that is relevant when the number of dimensions, $D$,
of the calculation is less than three.
It does not affect the steady state, but it affects the cooling rate both
in the optically thick and thin regimes; see \App{AppA} for details.
In one dimension with $D=1$, we have $N=2$ rays, which are
$\nnn_{1,2}=(0,0,\pm1)$, 
while for $D=2$ we can either have $N=4$ with $\nnn_{1,2}=(\pm1,0,0)$
and $\nnn_{3,4}=(0,0,\pm1)$, or $N=8$ with the additional 4 combinations
$\nnn_{5,...,8}=(\pm1,0,\pm1)/\sqrt{2}$.
In three dimensions, the correction factor is $D/3=1$, so the angular
integral is just $4\pi$ times the average of the intensity over all
directions.

\subsection{Parameters and initial conditions}
\label{Parameters}

In the following, we measure length in Mm, speed in $\km\s^{-1}$,
density in $\g\cm^{-3}$, and temperature in K.
\begin{table}[t]\caption{
Units used in this paper and conversion into cgs units.
}\vspace{12pt}\centerline{\begin{tabular}{lll}                            
\hline\hline
quantities&code units&cgs units\\
\hline                            
length [$z$]       &    $\Mm$    & $10^8\cm$\\
velocity {[$u$]}     &   $\km\s^{-1}$    & $10^5\cm\s^{-1} $ \\
density {[$\rho$]}&    $\g\cm^{-3} $ &$1\g\cm^{-3}$\\
temperature {[$T$]}     & $\K$ & $1\K$\\
time {[$t$]} & $\ks$ & $10^3\s$ \\
gravity {[$g$]}& $\km^2\s^{-2}\Mm^{-1}$& $10^2\cm\s^{-2}$\\
opacity  {[$\kappa$]}& $\Mm^{-1}\cm^3\g^{-1}$ &
$10^{-8}\cm^2\g^{-1}$\\
diffusivity [$\chi$] & $\Mm\km\s^{-1}$ & $10^{13}\cm^2\s^{-1}$\\
conductivity [$K$]$\!\!$ & $\g\cm^{-3}\km^3\s^{-3}\Mm\K^{-1}\!\!\!$ &
$10^{23}\g\cm\s^{-3}\K^{-1}\!\!\!$\\
Stefan-B [$\sigmaSB$] & $\g\cm^{-3}\km^3\s^{-3}\K^{-4}$ &
$10^{15}\g\s^{-3}\K^{-4}$\\
flux [$F$] & $\g\cm^{-3}\km^3\s^{-3}\!\!\!$ & $10^{15}\erg\cm^{-2}\s^{-1}$\\
\hline
\label{units}\end{tabular}}\end{table}
This implies that time, for example, is measured in $\ks$ (=$1000\s$).
The advantage of using this system of units is that it avoids extremely
large or small values of various quantities by using units that are
commonly used in solar physics such as Mm and km/s.
A summary of our units and the conversion of various quantities between
cgs and our units is given in \Tab{units}.

For the gravitational acceleration, we take $\grav=(0,0,-g)$ with
$g=274\km^2\s^{-2}\Mm^{-1}$ being the solar surface value \citep{Stix02}.
Instead of prescribing $T_{\rm bot}$, we prescribe the sound speed $\cs$,
where $\cs^2=\gamma{\cal R}T/\mu$, and fix $\cs=\csz=30\km\s^{-1}$
at $z_{\rm bot}=0$.
With ${\cal R}=8.314\;10^7\erg\K^{-1}\mol^{-1}$ and
$\mu=0.6\g\mol^{-1}$, this choice corresponds to $T_{\rm bot}=38,968\K$.
We found it instructive to start with an isothermal solution that
is in hydrostatic equilibrium, but not in thermal equilibrium, so
the upper parts will gradually cool until a static solution is reached.
Thus, we use $\rho=\rho_0\exp(-z/\Hp)$,
where $\Hp={\cal R}T/\mu g$ is the pressure scale height,
and $\rho_0$ is a constant that we set to
$\rho_0=4\;10^{-4}\,\g\,\cm^{-3}$.
This value was chosen based on values from a solar model
at a depth of approximately $7\Mm$ below the surface.
However, this particular choice is quite uncritical and just
corresponds to renormalizing the opacity.
In other words, instead of making a calculation with a ten times
larger value of $\rho$, we can just use an otherwise equivalent
calculation with a ten times larger value of $\kappa$.

\subsection{Simulation strategy}
\label{SimulationStrategy}

We choose the exponents $a$ and $b$ such that they correspond to
five different values of $n$.
In the case $a=-1$, $b=3$, we have $K(\rho,T)=\const$,
but the value of $n$ is undefined.
Our choice of parameters is summarized in \Tab{abn}.
\begin{table}[t!]\caption{
Summary of used $a$ and $b$.
}\vspace{12pt}\centerline{\begin{tabular}{crccl}
\hline \hline
Set & $a$  & $ b$  &  $n$ & Schwarzschild \\
\hline
A & 1 & $-3.5$ & 3.25&stable\\
B & 1 & 0 &1.5& marginally stable \\
C & 1 & 1 & 1 & unstable\\
D & 1 & 5 & $-1$& ultra unstable\\
E &$-1$ & 3 & $0/0$& undefined\\
\label{abn}\end{tabular}}
\tablefoot{
Combinations of exponents $a$ and $b$ and the resulting
polytropic index $n$ used in the present study.
The characterization with respect to the Schwarzschild stability
criterion is based on $\gamma=5/3$, corresponding to a marginal
polytropic index of $n=1/(\gamma-1)=3/2$.
Each parameter combination is denoted by a letter A--E,
which corresponds later to different sets of runs.
}
\end{table}
It is convenient to express $\kappa$ in the form
\EQ
\kappa=\tkapz\bigg({\rho \over \rho_0}\bigg)^a \bigg({T\over
  T_0}\bigg)^b,
\label{kappa-ab}
\EN
where $\tkapz$ is a rescaled opacity and is related to $\kappa_0$ by
$\tilde\kappa_0=\kappa_0\rho_0^a T_0^b$;
where $T_0=T_{\rm bot}$ is used.
(By contrast, $\rho_0$ is only approximately equal to the
density at the bottom---except initially.)
With this choice, the units of $\tkapz$ are independent of
$a$ and $b$, and always $\Mm^{-1}\cm^3\g^{-1}$
(=$10^{-8}\cm^2\g^{-1}$).
For each value of $n$, we choose 4 different values of $\tkapz=10^4$,
$10^5$, $10^6$, and $10^7\Mm^{-1}\cm^3\g^{-1}$.
We note that the actual Kramers opacity for free--free and bound--free
transitions with $a=1$ and $b=-7/2$ has $\kappa_0$ between
$6.6\;10^{22}$ and $4.5\;10^{24}\cm^5\g^{-2}\,K^{7/2}$,
respectively \citep{KW90}.
This corresponds to $\tkapz=2.26\;10^{11}$ and
$\tkapz=1.54\;10^{13}\Mm^{-1}\cm^3\g^{-1}$,
which are respectively four and six orders of magnitude larger than the largest
value considered in this paper.

\section{Results}

We perform one-dimensional simulations with a resolution of 512
equally spaced grid points using 
five sets of values for the exponents $a$ and $b$ in the expression for the
Kramers opacity; see \Eq{kappa-ab}.
Each set of runs is denoted by a letter A--E.
In the first four sets of runs, we keep $a=1$ and change the value of $b$ from
$-7/2$, to 0, 1, and 5.
For each of these sets, we perform four runs that differ only in the values
of $\tkapz$.  
The numeral on the label of each run refers to a different value of
$\tkapz$.
In Set~A, we use $a=1$ and $b=-7/2$.
Runs~A4, A5, A6 and A7 correspond to $\tkapz$
equal to $10^4$, $10^5$, $10^6$ and $10^7\,\Mm^{-1}\cm^3\g^{-1}$, respectively. 
All the other designations follow the same scheme.
All runs have been started with the same isothermal initial condition.
However, the size of the domain is changed so as to accommodate the upper
isothermal part by a good margin.
If the domain is too big, one needs a large number of meshpoints to resolve
the resulting strong stratification, and if it is too small, the solution
changes in the top part, as will be discussed in \Sec{size}.

\begin{figure}[t!]
\begin{center}
\includegraphics[width=\columnwidth]{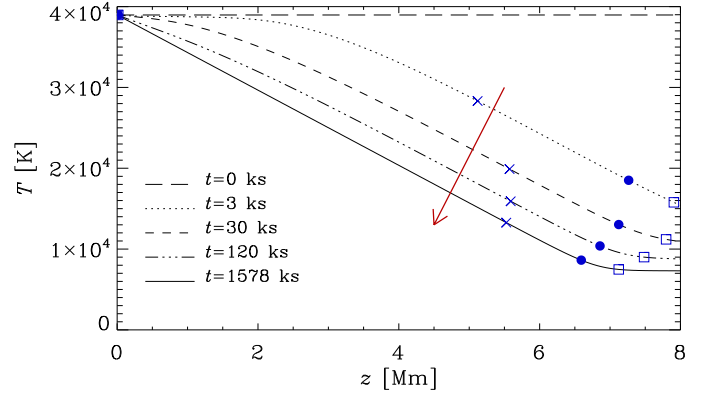}
\end{center}\caption[]{Vertical temperature profile at five different times
$t=$ 0, 3, 30, 120, and $1578\ks$ for Run~A6 with 
$\tkapz=10^6\,\Mm^{-1}\cm^3\g^{-1}$. 
Squares, circles and crosses represent different optical depths
$\tau=0.1$, $\tau=1$ and $\tau=10$, respectively.
The arrow represents the direction of the time evolution of the
temperature profile. 
}\label{tt-evol}\end{figure} 

After a sufficient amount of running time, a unique equilibrium state 
is reached and the resulting profiles of temperature,
density and entropy have a nearly polytropic stratification in the lower part
of the domain and a nearly isothermal stratification in the upper part
of the domain. 
An exception are the runs of Set~E where the polytropic index is undefined
($n=0/0$).
This will be discussed in more detail in \Sec{sec:n00}.
We summarize the important quantities obtained from all runs in
\Tab{results}.
These quantities are calculated in the equilibrium state.
All runs show a similar evolution of density, temperature and entropy.
In the next sections we describe the resulting profiles in more detail.

\begin{table*}[t!]\caption{Summary of the runs.
}\vspace{12pt}\centerline{\begin{tabular}{ccccccccrrcc}
\hline \hline
Run&$a$&$b$&$n$ &$\tkapz$ & $z_{\rm top}$ & $z_{\tau=1}$ & $\rho_{\tau=1}$ & $\tau_{\rm adjust}$ & $\Teff$ & $K_{\rm bot}$\\
\hline
A4&1&$-3.5$&3.25 &$10^4$ & 8 & 2.8 & $1.0\;10^{-4}$ &  50 & 23600 &$3.9\;10^{-6}$ & \\ 
A5&1&$-3.5$&3.25& $10^5$ & 8 & 5.2 & $1.7\;10^{-5}$ &  90 & 13900 &$4.6\;10^{-7}$ & \\ 
A6&1&$-3.5$&3.25& $10^6$ & 8 & 6.6 & $2.5\;10^{-6}$ & 700 &  7800 &$4.6\;10^{-8}$ & \\ 
A7&1&$-3.5$&3.25& $10^7$ & 8 & 7.4 & $3.7\;10^{-7}$ & 15000 &  4400 &$4.4\;10^{-9}$\\ 
\hline
B4&1&0&1.5&$10^4$ & 5 & 1.4 & $2.2\;10^{-4}$ & 40 & 26600 &$4.53\;10^{-6}$ \\ 
B5&1&0&1.5&$10^5$ & 5 & 2.9 & $9.0\;10^{-5}$ & 60 & 16300 &$5.15\;10^{-7}$ \\ 
B6&1&0&1.5&$10^6$ & 5 & 3.8 & $3.7\;10^{-5}$ & 400 &  9300 &$5.38\;10^{-8}$ \\ 
B7&1&0&1.5&$10^7$ & 5 & 4.3 & $1.6\;10^{-5}$ & 5000 &  5200 &$5.08\;10^{-9}$ \\ 
\hline
C4&1&1&1& $10^4$ & 4 &  1  & $2.6\;10^{-4}$ & 7 & 27600 &$5.1\;10^{-6}$  \\ 
C5&1&1&1& $10^5$ & 4 & 2.3 & $1.3\;10^{-4}$ & 20 & 17400 &$5.6\;10^{-7}$  \\ 
C6&1&1&1& $10^6$ & 4 & 3.1 & $7.0\;10^{-5}$ & 200  & 10100 &$6.0\;10^{-8}$  \\ 
C7&1&1&1& $10^7$ & 4 & 3.4 & $3.9\;10^{-5}$ & 2100 & 5700  &$6.1\;10^{-9}$  \\ 
\hline
D4&1&5&$-1$& $10^4$ & 2 & 0.2 & $3.6\;10^{-4}$ & 6 & 31000 &$1.1\;10^{-5}$  \\ 
D5&1&5&$-1$& $10^5$ & 2 & 0.8 & $2.8\;10^{-4}$ & 7 & 23100 &$1.3\;10^{-6}$  \\ 
D6&1&5&$-1$& $10^6$ & 2 &  1  & $2.8\;10^{-4}$ &  80  & 15600 &$1.9\;10^{-7}$  \\ 
D7&1&5&$-1$& $10^7$ & 2 &  1  & $3.2\;10^{-4}$ & 700  & 10100 &$3.1\;10^{-8}$  \\ 
\hline
E4&$-1$&3&0/0& $10^4$ & 4 & 3.0 & $8.7\;10^{-5}$ & 6 & 23700 &$4.47\;10^{-6}$ &  \\ 
E5&$-1$&3&0/0& $10^5$ & 4 & 3.6 & $5.6\;10^{-5}$ & 45 & 14900 &$4.47\;10^{-7}$ &  \\ 
E6&$-1$&3&0/0& $10^6$ & 4 & 3.8 & $3.9\;10^{-5}$ & 400 & 8800 &$4.47\;10^{-8}$ &  \\ 
\label{results}\end{tabular}}
\tablefoot{
The size of the domain $z_{\rm top}$ and the
height of the surface $z_{\tau=1}$ are in Mm, the
density at the surface $\rho_{\tau=1}$ is in $\g\cm^{-3}$,
the thermal adjustment time $\tau_{\rm adjust}$ is in $\ks$,
the effective temperature $T_{\rm eff}$ is in $\K$,
radiative heat conductivity at the bottom of the domain $K_{\rm bot}$
is in $\g\cm^{-3}\km^3\s^{-3}\Mm\K^{-1}$ and the normalized
opacity $\tkapz=\kappa_0\rho_0^a T_0^b$
is in $\Mm^{-1}\cm^3\g^{-1}$ are shown for each run.
The second to the sixth columns show quantities which are input
parameters to the models whereas the quantities in last five columns
are the results of the simulations, computed from the equilibrium
state.
}
\end{table*}

\subsection{Approach toward the final state}
\label{Approach}

\begin{figure}
\begin{center}
\includegraphics[width=\columnwidth]{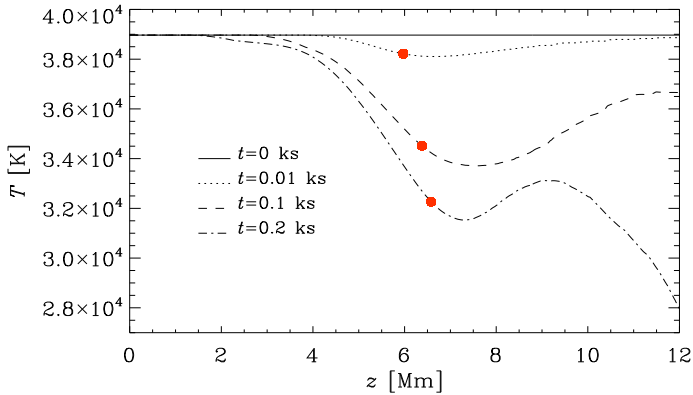}
\end{center}\caption[]{Temperature over height for Run~A6 at different times
  $t=0$, $0.01$, $0.1$, and $0.2\,$ks, plotted as solid,
 dotted, dashed and dashed-dotted lines, respectively. The
 red dots correspond to the location of $\tau\approx 1$.  
}\label{cooling}\end{figure}

As mentioned above, we find it convenient to obtain equilibrium solutions
by starting from an isothermal state.
The upper layers begin to cool fastest, and eventually an equilibrium
state is reached.
We plot the evolution of the temperature profile of Run~A6 in
\Fig{tt-evol} as an exemplary case with
$\tkapz=10^6\,\Mm^{-1}\cm^3\g^{-1}$.
Already after a short time of $t=3\ks$ (1\,hours), the temperature has
decreased by more than half its initial value at the top and
follows a polytropic solution in most of the domain, where the temperature
gradient has a similar value than in the equilibrium state.
At $t=30\,$ks (8\,hours), close to the top boundary, an isothermal
part is seen to emerge. However, it takes more than $t=1500\,$ks
(17\,days) until the equilibrium 
solution is reached with a prominent isothermal part of $T\approx7000\,$ K.
The locations of three different optical depths,
$\tau=0.1,\,1$ and $10$, are shown in \Fig{tt-evol}. 
Here,
\EQ
\tau(z)=\int_z^{z_{\rm top}}\kappa(z')\rho(z')\,\dd z'
\EN
is the optical depth with respect to the surface of the domain.
If the domain is tall enough, one can see that an initially
isothermal stratification cools down first 
near the location where $\tau(z)\approx 1$,
which is where the cooling is most efficient. 
As an example we plot in \Fig{cooling} the early stages of the temperature
evolution at $t=0$, 0.01, 0.1, and $0.2\ks$ for a taller variant of
Run~A6 with $z_{\rm top}=12\Mm$ using 1024 equally spaced grid points.
At $t=0.01\ks$, the temperature starts to
decrease at the height where $\tau\approx1$, while in the upper part,
which is far enough from the surface, the temperature remains at first
unchanged. 
Only at a somewhat later time ($t=0.1\ks$) does the temperature at $z=12\Mm$
start to cool down.
This is explained by the fact that the radiative cooling rate
(or inverse cooling time)
is largest near $\tau=1$ \citep{Spi57,US66,Edw90}; see also \App{AppA}.

\subsection{Temperature stratification}
\label{TDS}

For all runs, the temperature reaches an equilibrium state
after a certain time; see \Fig{tt-evol}.
The temperature profile can be divided into two distinguishable
parts,
a nearly polytropic part which starts from the bottom of the domain
and extends to
a certain height, and a nearly isothermal part which starts from
this height and extends to the top of the domain.
The transition of the temperature from the initial state to the
equilibrium state follows a specific pattern, which is the same for all
the runs.
The higher the value of $\tkapz$, the lower the temperature is in the
isothermal part and the longer it takes to reach this state.
Increasing the normalized opacity $\tkapz$ by three orders of magnitude
results in a decrease in the temperature
by a factor of five for Set~A and a
factor of three for Set~D.
As the exponent $b$ changes from the smallest value in
Set~A to the largest one in Set~D, the slope of the temperature decreases
with height. This means that the polytropic part of the atmosphere is
smaller for larger values of $b$.
We note that the size of the domain is chosen larger for smaller values
of $b$.
For Sets~A, B and C, the temperature in the polytropic part is almost
the same for different values of $\tkapz$,
although for the lowest value of $\tkapz$ the temperature deviates somewhat.
However, in Set~D, for different values of $\tkapz$, the slope of the
temperature is different for each value of $\tkapz$.
This has to do with the fact that in this case with $(b-3)/(1+a)=-1$ the
stratification is no longer a polytropic one.
(A polytrope with $n=-1$ would have constant pressure, which is unphysical.)

The temperatures in the isothermal part also show a dependency on $b$.
For $\tkapz=10^4\,\Mm^{-1}\cm^3\g^{-1}$ the temperature in Run~A4 is
$T\approx2.2\;10^4\,$K, whereas in Run~D4 the value is
$T\approx2.9\;10^4\,$K. 
A similar behavior can also be seen for the other values of $\tkapz$.
\begin{figure*}[t!]
\begin{center}
\includegraphics[width=2\columnwidth]{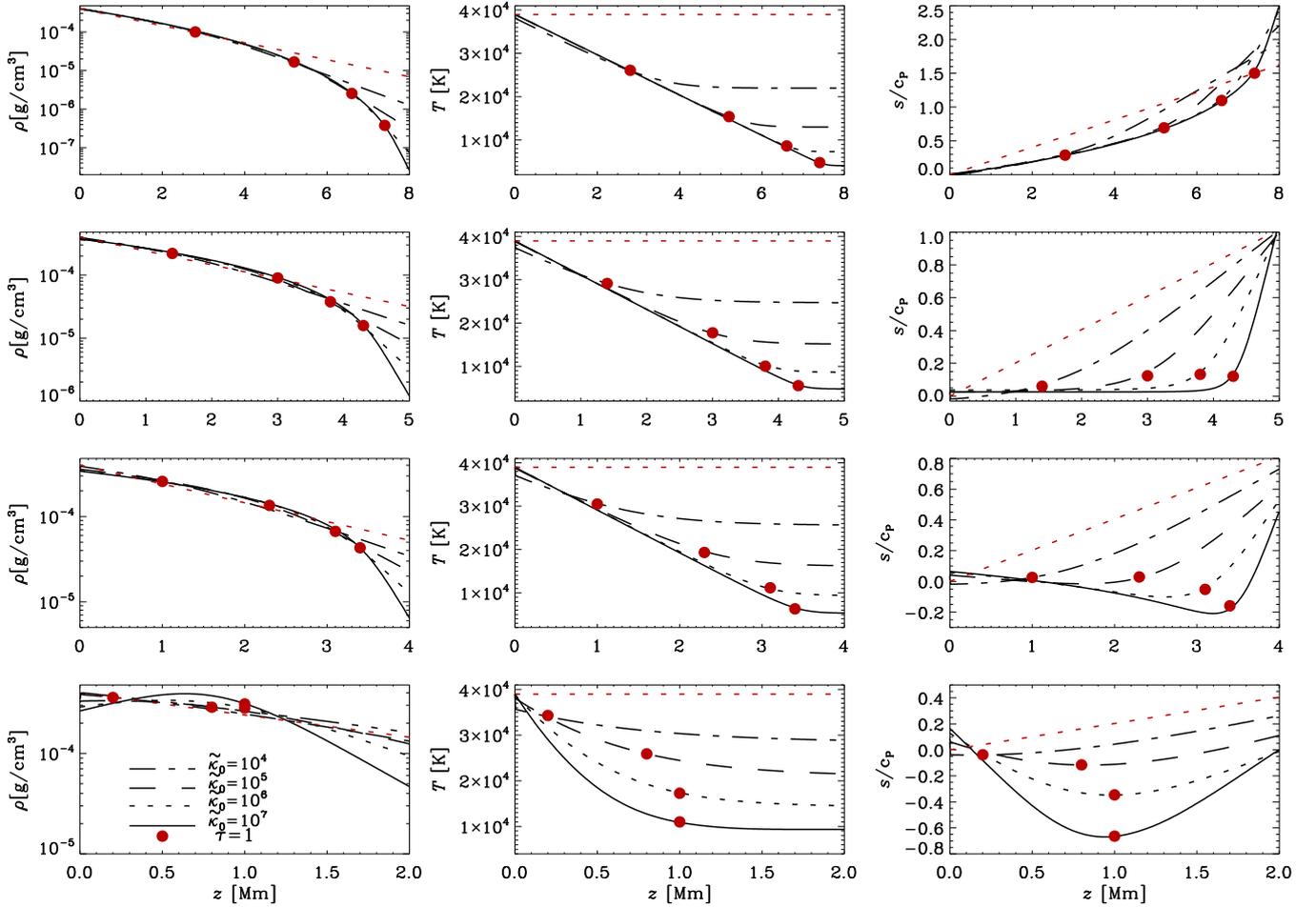}
\end{center}\caption[]{Density, temperature and entropy of the
  equilibrium state versus height,
from left to right, for Sets~A, B, C and D,
  from top to bottom. 
The four different lines in each plot corresponds to the value of the
rescaled opacity $\tkapz=10^4$, 
$10^5$, $10^6$, $10^7\,\Mm^{-1}\cm^3\g^{-1}$.
The dots in all plots represent the surface $\tau\approx1$.
The red dotted lines represent the initial profile of each set.     
}\label{comp}\end{figure*}
Next, we calculate the optical depth for all runs.
We find that the transition point from the polytropic part to
the isothermal part coincides with the $\tau\approx 1$ surface.
We indicate the surface $\tau\approx 1$ by dots in all plots
in \Fig{comp}.
The polytropic part corresponds to the optically
thick part with $\tau>1$ and the isothermal part corresponds to the
optically thin part with $\tau<1$. For each set, the transition point
depends on the value of $\tkapz$.
As we go from smaller to larger values of $\tkapz$,
the surface shifts to larger heights and becomes cooler.
This is because the radiative heat conductivity $K$ is
inversely proportional to $\tkapz$ and directly proportional to the flux.
Therefore, by increasing the value of $\tkapz$, $K$ decreases and,
as a consequence, the radiative flux also decreases.
By decreasing the flux, the effective temperature decreases as 
$T_{\rm eff}\propto F_{\rm rad}^{1/4}$.
This means that the temperature at the surface is smaller for
larger values of $\tkapz$. 
For Set~A, the $\tau\approx 1$ surfaces lie on the polytropic
part of the temperature profile.
However, by increasing the value of $b$, the locations of the $\tau=1$ surfaces
shift toward the lower boundary and the optically thick part becomes narrower.
This is particularly severe for the solutions with
a small value of $\tkapz$, especially for $\tkapz=10^4\,\Mm^{-1}\cm^3\g^{-1}$,
when the boundary condition
$T=T_{\rm bot}$ at $z=0$ becomes unphysical and the temperature drops between
the first two meshpoints in a discontinuous fashion; see \Fig{comp}.

\subsection{Entropy stratification}
\label{strat-ss}
We plot the entropy profiles for all sets of runs in the equilibrium state
in the last column of \Fig{comp}.
For Runs~C6--7 and D5--7, the entropy decreases in the polytropic
part and starts to increase in the isothermal part.
All runs show a positive vertical entropy gradient in the isothermal part.
In the lower part, the entropy gradient is positive
($\nabla_z s>0$) for Set~A, while for Set~B it is constant and 
equal to zero ($\nabla_z s\approx 0$).
This shows that for Set~B, the atmospheres are isentropic. 
In Sets~C and D, except for the case
$\tkapz=10^4\,\Mm^{-1}\cm^3\g^{-1}$, the 
entropy gradient is negative, $\nabla_z s<0$.
This means that their atmospheres are convectively unstable. 
(Convection will of course not occur in our one-dimensional model, but we
will obtain the so-called hydrostatic reference solution that is used to
compute the Rayleigh number, as will also be done later in this paper.)
In Set~D the entropy gradients are larger than in
case C where their atmospheres are marginally stable.
In the isothermal part of Set~C, the entropy gradient is much
larger than in Set~D. 
For each set of runs, as we go from smaller values of $\tkapz$ to larger
ones, the entropy profiles have larger gradients.

\subsection{Incoming and outgoing intensity profiles}
\label{RT-cool}

\begin{figure}[t!]
\begin{center}
\includegraphics[width=\columnwidth]{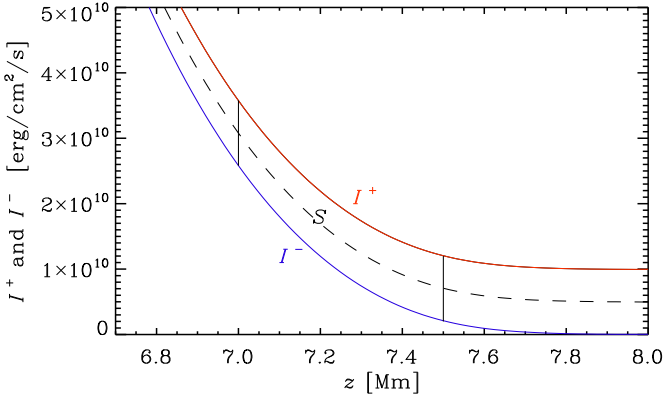}
\end{center}\caption[]{
Source function and vertical profiles of
incoming and outgoing
intensity near the surface for Run~A7.
The dashed line represents the source function and the solid lines
represent the incoming intensity $I^+$ (red) and outgoing
intensity $I^-$ (blue). 
The vertical lines represent the (constant) difference between $I^+$ and $I^-$.
}\label{diff}\end{figure}

It is instructive to inspect the vertical profiles of the intensity
for rays pointing in the up- and downward directions, $\nnn=(0,0,\pm1)$,
denoted in the following by $I^\pm$.
If we have just these two rays, the energy flux is given by
$F_{\rm rad}=(2\pi/3)(I^+-I^-)$.
In thermal equilibrium, the difference between $I^+$ and $I^-$
must be constant.
This is indeed the case; see \Fig{diff}, where we plot $I^+$ and $I^-$
and compare with $S$.
The vertical lines in this figure represents the difference between $I^+$ and
$I^-$, where $I^+-I^-\approx10^4\,$
$\erg\cm^{-2}\s^{-1}$ in the whole
domain as we have radiative equilibrium $\nab\cdot\FF_{\rm rad}=0$. 
Radiative equilibrium also demands that $J=S$, so $S$ has to be equal
to the average of $I^+$ and $I^-$, which is indeed the case.

\subsection{Radiative heat conductivity}
\begin{figure}[t!]\begin{center}
\includegraphics[width=\columnwidth]{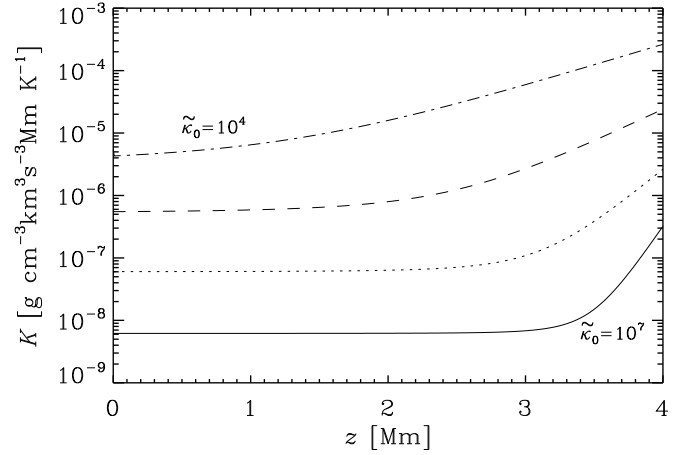}
\end{center}\caption[]{
    Radiative heat conductivity $K$ versus height for Set~C.
    $K$ is plotted for different values of $\tkapz$
 where $\tkapz=10^4\,\Mm^{-1}\cm^3\g^{-1}$ is shown by
    dotted-dashed line, $\tkapz=10^5\,\Mm^{-1}\cm^3\g^{-1}$ dashed
    line, $\tkapz=10^6\,\Mm^{-1}\cm^3\g^{-1}$ dotted line and 
    $\tkapz=10^7\,\Mm^{-1}\cm^3\g^{-1}$ solid line.
}\label{pK}\end{figure}

In Sets~A, B, and C, the value of radiative heat conductivity $K$
turns out to be 
constant in the optically thick part of the atmosphere, but not for
Set~D. 
The value of $K$ at the bottom of the optically thick part of the domain
is denoted in \Tab{results} by $K_{\rm bot}$, and agrees roughly with
\EQ
K_{\rm bot}\approx K_0\equiv{16 \sigmaSB T_0^3\over 3\tkapz\rho_0}.
\label{Kbot}
\EN
Indeed, it has almost the same order of magnitude for A, B and C,
independently of the value of $b$, but it is one order of magnitude
larger for Set~D.  
This is because in this set the density is lower
in the optically 
thick part compared to the other sets.
Moreover, as we go from higher values of $\tkapz$ to the lower ones, the
radiative heat conductivity increases.
This can be explained by the inverse proportionality of $K$ with opacity as
$K\propto 1/\tkapz$.
For smaller values of $\tkapz$, $K$ is larger and vice versa.   
As an example, we plot in \Fig{pK}
the resulting vertical profiles of radiative
heat conductivity for Set~C.
We note that $K$ is constant in the optically thick part and
starts to increase in the optically thin part.
In the optically thin part, $\kappa\rho$ decreases, so
$K$ increases as $K\propto 1/\kappa\rho$.
To maintain $\nab\cdot \FF_{\rm rad}=0$, the modulus of $\nab T$ has to decrease.
As $K$ increases even further, a thermostatic equilibrium can be
satisfied if $\nab T$ comes close to zero.

\subsection{Effective temperature}

\begin{figure}[t!]\begin{center}
\includegraphics[width=\columnwidth]{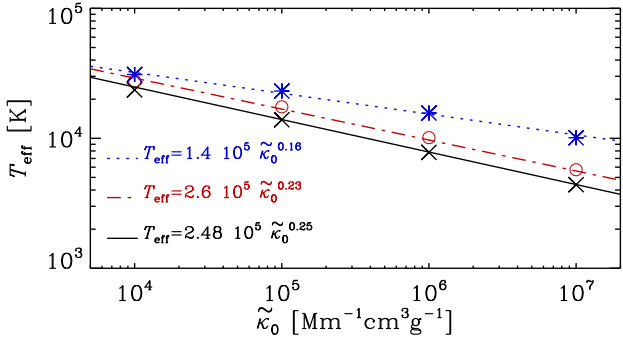}
\end{center}\caption[]{
Effective temperature $T_{\rm eff}$ versus rescaled opacity $\tkapz$ for 
Sets~A, C, and D.
The crosses, circles and stars show the values of $T_{\rm eff}$ for
different values of $\tkapz$ for Sets~A, C, and D, respectively.
Different lines correspond to line fit of $T_{\rm eff}$ with
normalized opacity $\tkapz$.
}\label{Teff-comp}\end{figure}

The effective temperature $T_{\rm eff}$ of all runs is calculated from the
$z$ component of the radiative flux $\FF_{\rm rad}$,
\EQ
T_{\rm eff}=\bigg({F_{\rm rad}\over \sigmaSB}\bigg)^{1/4}.
\EN
The values of $T_{\rm eff }$ of all sets of runs are summarized in \Tab{results}.
By increasing the value of $b$, $T_{\rm eff }$ also increases.
The value of $T_{\rm eff }$ decreases as we go from lower to higher
opacities for each set.
We plot $T_{\rm eff}$ versus $\tkapz$ in \Fig{Teff-comp} for Sets~A, C
and D where the values of $T_{\rm eff}$ are represented by crosses,
circles and stars, respectively. 
For each set of runs, we fit a line to $T_{\rm eff}$ versus $\tkapz$. 
We find that $T_{\rm eff}$ has a power law relation with $\tkapz$.
The power of $\tkapz$, which is the slope of the plot, depends on the
polytropic index and therefore on $b$.
For larger values of $b$, the power is smaller than for smaller values of $b$.
Additionally, the offset shows also a weak dependence on $b$.
A power law relation between $T_{\rm eff}$ and the opacity of roughly
$1/4$ can be expected, because of the linear relation of the radiative
flux and the opacity.
Toward larger $b$, this dependency is no longer accurate.
We also calculate for each run the corresponding optical depth
where $T=T_{\rm eff}$.
For all runs, $T_{\rm eff}$ corresponds to the optical depth
$\tau\approx 1/3$.
This is less than what is expected for a gray
atmosphere, where $T_{\rm eff}=T$ at $\tau\approx 2/3$. 
This is presumably because in our integration of $\tau$ we have not
included the contribution between $\infty$ and $z=z_{\rm top}$. 

\subsection{Thermal adjustment time}
\label{t-adjust}

\begin{figure}[t!]
\begin{center}
\includegraphics[width=\columnwidth]{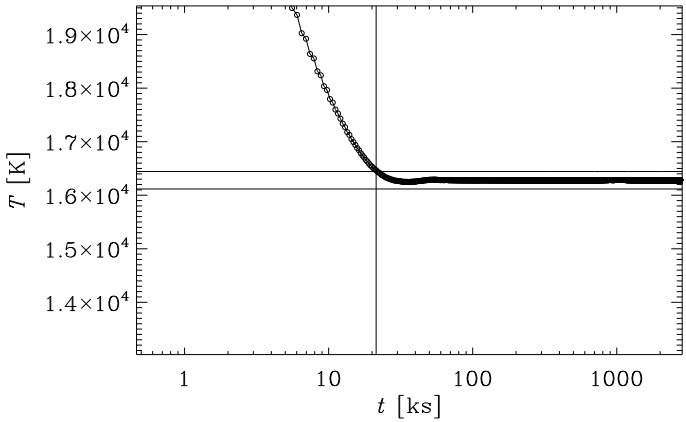}
\end{center}\caption[]{
Temperature $T$ at $z_{\rm top}=4\Mm$ versus time $t$ for Run~C5.
The two horizontal lines mark the 1\% margin around the final value of $T$
and the vertical line marks the time $\tau_{\rm adjust}\approx20\ks$
after which $T$ lies within these margins.
}\label{Txy}\end{figure}

In our simulations we define a thermal adjustment time $\tau_{\rm adjust}$
as the time it takes for each run to reach its numerically obtained
final equilibrium temperature in the isothermal part to within $1\%$
(see \Fig{Txy}). 
The unit of $\tau_{\rm adjust}$ is ks (see \Sec{Parameters}).
The value of $\tau_{\rm adjust}$ for all runs is summarized in \Tab{results}.
As we can see in \Tab{results}, the thermal adjustment time becomes
smaller for larger $b$ and smaller $n$.
For each set of runs, $\tau_{\rm adjust}$ grows approximately linearly with
$\tkapz$, although for $\tkapz\la10^5\Mm^{-1}\cm^3\g^{-1}$,
the dependency is more shallow.
For larger values of $\tkapz$, $\tau_{\rm adjust}$ seems to have a
stronger dependency on $b$.
We speculate that the reason for increasing the value of $\tau_{\rm adjust}$
for higher values of $\tkapz$ is that by increasing the opacity the
energy transport 
via radiation becomes less efficient as the mean free path of the
photon decreases.
But it seems that there exists a threshold of efficiency, leading to a larger
adjustment time for the lowest values of $\tkapz$, as expected.

We plot the vertical dependence of the mean free path of the photons
$\ell=1/\kappa\rho$ normalized by the size of the domain $L$ for Set~C.
\begin{figure}[t!]\begin{center}
\includegraphics[width=\columnwidth]{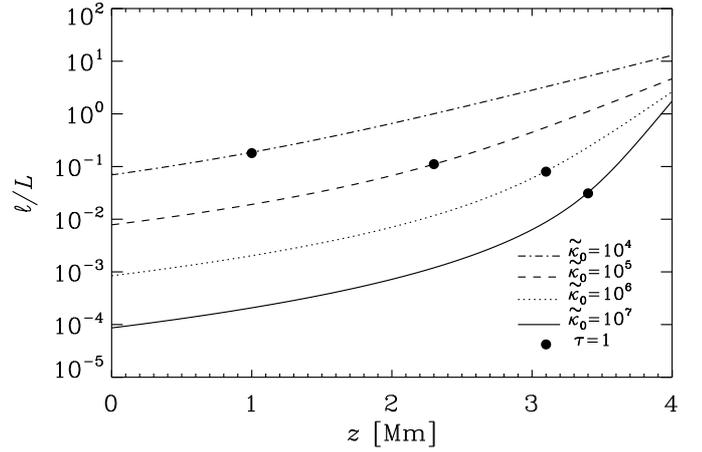}
\end{center}\caption[]{Normalized mean free path of photons
$\ell/L$ versus height for Set~C.
    The dots represent the surface $\tau\approx1$.
}\label{lambda-photon}\end{figure}
As we can see in \Fig{lambda-photon}, the mean free path increases
by several orders of magnitude from the bottom of the domain to the top.
Furthermore, $\ell$ is larger for smaller $\tkapz$.
In the optically thick part, the difference in $\ell$ is one order
of magnitude, which is equal to a corresponding change in $\tkapz$.
In the optically thin part, the difference in the values of
$\ell$ becomes smaller, as we reach the top of the domain.
For $\tkapz=10^7\,\Mm^{-1}\cm^3\g^{-1}$, $\ell$ is the
smallest and,
at the bottom of the domain,
three orders of magnitude smaller than for
$\tkapz=10^4\,\Mm^{-1}\cm^3\g^{-1}$.
Furthermore, $\ell$ is 10 times the size of the domain for
$\tkapz=10^4\,\Mm^{-1}\cm^3\g^{-1}$,
which makes the cooling more efficient.
We would have expected to see a large change in the mean free path as
we go through the surface.
Nevertheless, the exponential growth seems to be roughly the same throughout the
domain, at least for the smallest value of $\tkapz$. 
 
\subsection{Properties of an atmosphere with undefined $n$}
\label{sec:n00}

\begin{figure}[t!]
\begin{center}
\includegraphics[width=\columnwidth]{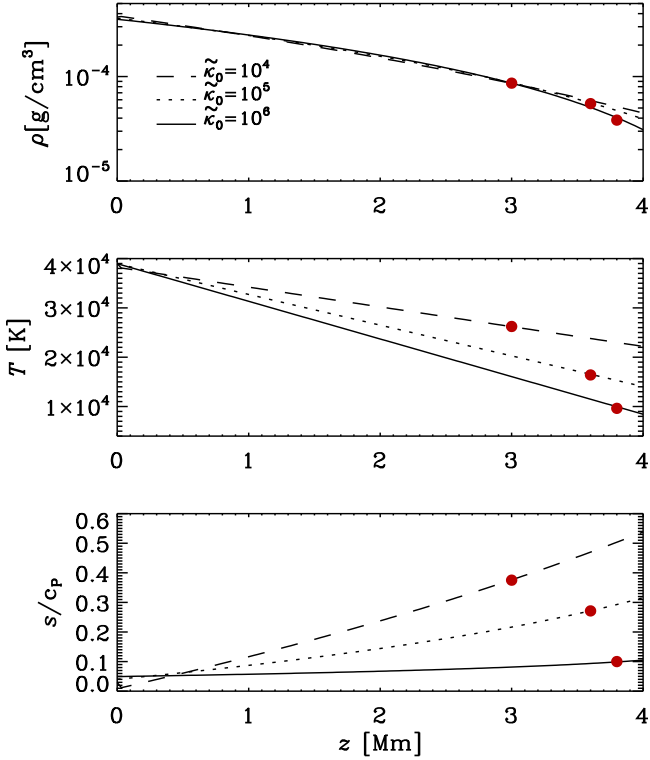}
\end{center}\caption[]{
Density, temperature, and entropy profiles for Set~E ($n=0/0$).
Dashed, dotted, and solid lines represent E4, E5, and E6, respectively.
The red dots present the surface of the model where $\tau=1$.
}\label{n00}\end{figure}

By choosing $a=-1$ and $b=3$, we have a constant heat conductivity $K$
that is independent of density and temperature as the heat
conductivity is given by \Eq{K-model}.
The nominal value of $n$ is given by 
\EQ
n={3-3\over1-1}={0\over 0}.
\EN
In this case, since $K=\const$,
we expect to have only a polytropic solution which
satisfies the thermostatic equilibrium if $\nabla_z T=$ const,
but it is then unclear how $\rho$ varies.
In \Fig{n00}, we plot the profiles of density, temperature and entropy
for all three runs of Set~E.
The first panel shows that in the optically thick part 
the density is nearly the same for the three values of $\tkapz$,
while in the optically thin part it decreases with increasing $\tkapz$.
For higher values of $\tkapz$, the density drops faster than in the case
of smaller $\tkapz$.
In all cases, $K$ is constant in both the optically thick and 
thin parts, but an interesting aspect is that its bottom value $K_{\rm bot}$ is
of the same order of magnitude as in Sets~A, B, and C.
In the second panel of \Fig{n00}, we plot the temperature profiles.
 As expected, there is no isothermal part. 
The slope of temperature decreases approximately linearly as we go to
higher values of 
$\tkapz$, because $K$ is proportional to $1/\tkapz$.
Although the solutions show no transition from
the polytropic part to an isothermal one, the atmosphere has still
a layer where $\tau=1$, which is shown as red dots in all panels of
\Fig{n00}.
In contrast to the other sets, A, B, C, and D, the temperature profiles
look qualitatively different.
As in Sets~A, B, and C, in the optically thick part, the different
temperature profiles have nearly the same gradient, while in Set~E, the
gradient is different for the three values of $\tkapz$.
This is because in this case, thermostatic equilibrium
is obeyed with a constant value of $K$ (independent of $z$).

In the third panel of \Fig{n00}, we plot entropy profiles for the
three values of $\tkapz$.
In all cases the entropy increases with a slope that
depends on $\tkapz$.
 \begin{figure}[t!]
\begin{center}
\includegraphics[width=\columnwidth]{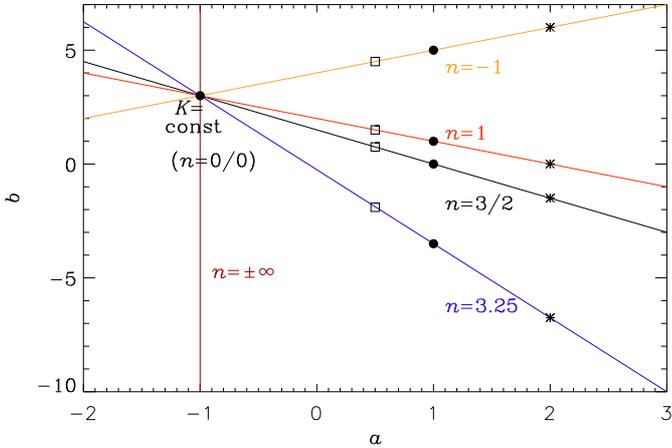}
\end{center}\caption[]{$b$ versus $a$ for different values of the
  polytropic index $n$. The black dots represents the combination of
  $a=1$ with different values of $b$ which are used in the main
  simulation; see \Tab{results}. The stars represent the combination of
  $a=2$ and squares represent $a=0.5$ with different values of $b$;
  see \Tab{test-n}.
}\label{ab-n}\end{figure}
The actual polytropic index can be computed from the resulting
super-adiabatic (or entropy) gradient,
\EQ
\nabla-\nabla_{\rm ad}={\dd(s/\cp)\over\dd\ln p},
\EN
where $\nabla_{\rm ad}=1-1/\gamma$ is the adiabatic gradient.
This gives $\nabla$, which is related to the actual $n$ via
\EQ
n_{\rm actual}={\dd\ln\rho\over\dd\ln T}
={\dd\ln p\over\dd\ln T}-1=\nabla^{-1}-1,
\label{nactual}
\EN
which follows from the perfect gas relation $p\propto\rho T$.
We note also that convection is not possible in one dimension, so we
obtain directly the hydrostatic solution, which may be unstable.

By solving \Eq{n_from_ab} for $b$, we obtain $b=3-n(1+a)$; see \Fig{ab-n}.
We see that for different values of $n$, the graphs of $b$ versus $a$
intersect each other at one common point.
This corresponds to $K=K_0=\const$; see \Eq{K-model}.
This means that the solution for constant $K$ can belong to any of
these polytropic indexes.
For $a=-1$ and $b=3$, in which case $n$ is undefined,
the solutions have a value of $n_{\rm actual}$ that
depends on the height of the domain.
Using \Eq{nactual} together with hydrostatic equilibrium,
$\dd p/\dd z=-\rho g$, we find
\EQ
n_{\rm actual}=g\,{\mu\over{\cal R}}\,
{z_{\rm top}-z_{\rm bot}\over T_{\rm bot}-T_{\rm top}}-1,
\label{nactual2}
\EN
so $n_{\rm actual}$ increases as $z_{\rm top}$ is increased.
However, this increase is partially being compensated by a small
simultaneous decrease of $T_{\rm top}$, which reduces the increase
of $n_{\rm actual}$ by about 10\% as $z_{\rm top}$ increases.
When the domain is sufficiently thin, the value of $n_{\rm actual}$
drops below the critical value $(\gamma-1)^{-1}=3/2$,
so the system would be unstable to the onset of convection.
We return to the relation between $n$ and the height of the domain in
\Sec{OpticallyThick}, where we consider solutions using the optically
thick approximation with a radiative boundary condition at the top.

\subsection{Dependence on the size of the domain}
\label{size}

\begin{figure}[t!]\begin{center}
\includegraphics[width=\columnwidth]{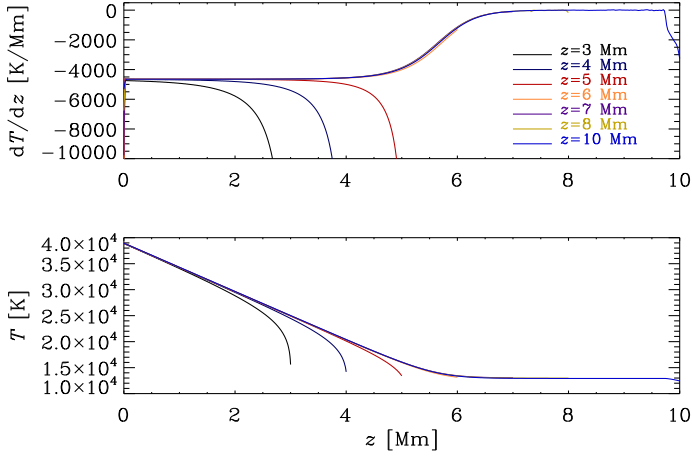}
\end{center}\caption[]{
Temperature gradient (upper panel) and temperature profile (lower
panel) of seven different sizes of the domain $z=3$, 4, 5, 6, 7, 8,
and $10\Mm$ of Run~A5.
We note that the lines for $z\ge6$ all fall on top of each other.
}\label{gradT}\end{figure}

In our model, the size of the domain plays an important role in getting
the polytropic and isothermal solutions for the temperature
profile.
The domain has to be big enough so that the transition point lies inside
the domain.
In \Fig{gradT}, we show the vertical dependence of temperature for six
domain sizes for Run~A5.
If the size of the domain is $z<7\Mm$, it is too small to obtain the
isothermal part where $\nabla_z T=0$ and a boundary layer is produced.
The opacity is then too large to let the heat
be radiated away.
A size of around $z=8\Mm$ is sufficient to get the isothermal part.
However, a domain size that is too large ($z=10\Mm$) leads to numerical
difficulties near the top boundary, especially if the resolution is too low.
For all the runs shown in \Tab{results}, we have always started by
performing several test simulations to find a suitable domain size.

\subsection{Radiative diffusivity}

\begin{figure}[t!]\begin{center}
\includegraphics[width=\columnwidth]{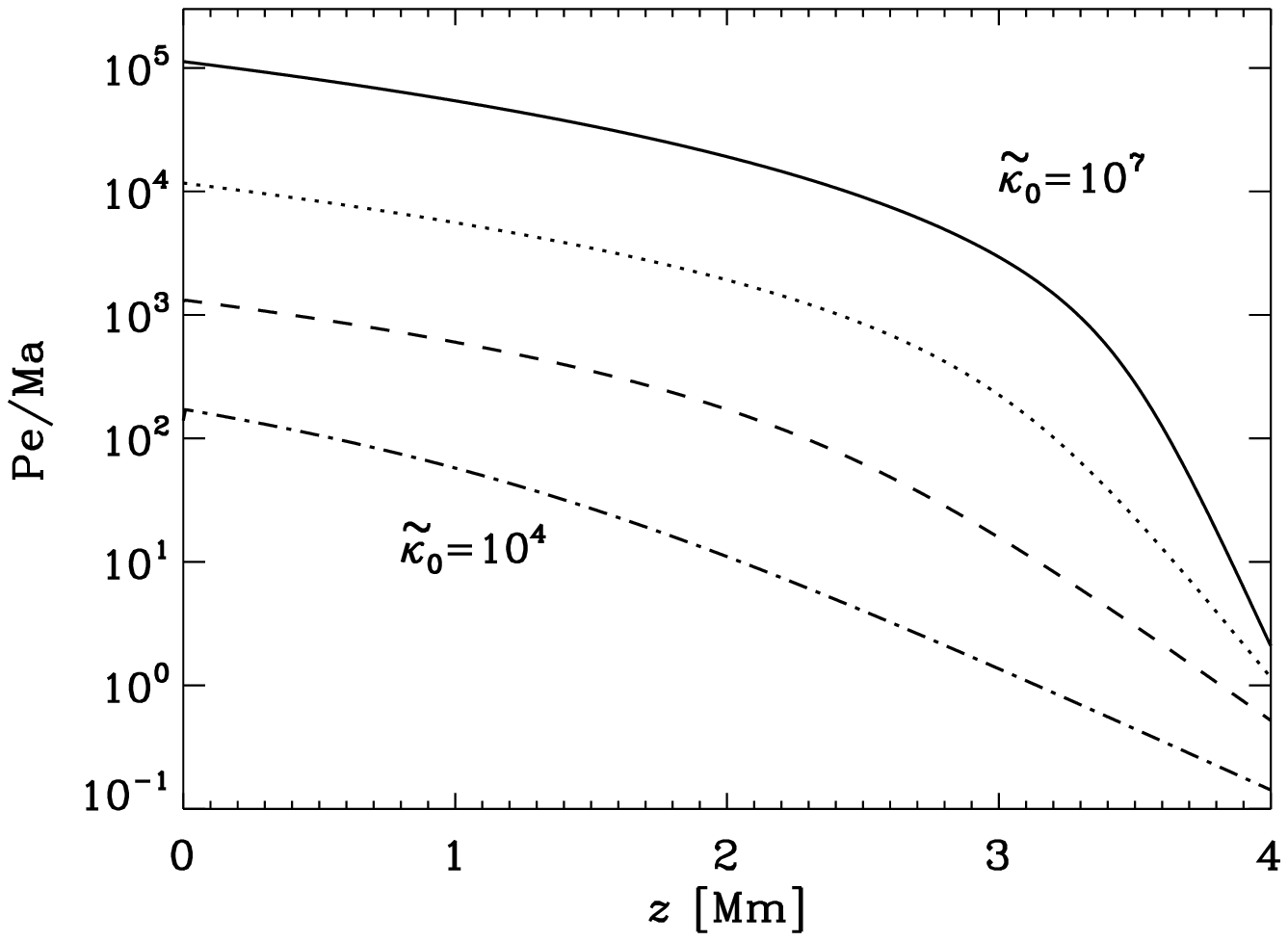}
\end{center}\caption[]{
$\Pet$ versus $z$ for Set~C using different values of $\tkapz$:
    $\tkapz=10^4\,\Mm^{-1}\cm^3\g^{-1}$ (dotted-dashed line),  
    $\tkapz=10^5\,\Mm^{-1}\cm^3\g^{-1}$ (dashed line),
    $\tkapz=10^6\,\Mm^{-1}\cm^3\g^{-1}$ (dotted line) and 
    $\tkapz=10^7\,\Mm^{-1}\cm^3\g^{-1}$ (solid line).  
}\label{Ra-Pe}\end{figure}

In numerical simulations, the radiative diffusivity $\chi$
is an important parameter, and has the same dimension as
the kinematic viscosity $\nu$.
Both $\chi$ and $\nu$ determine whether the results of numerical turbulence
simulations are reliable or not and whether they are able to dissipate
all the energy within the mesh. 
In a numerical simulation we are restricted to a certain number of
grid points.
If the diffusion of the temperature in a simulation is very small, 
it can happen that the changes in the temperature are too large over the
distance of neighboring grid points.
Hence, the changes in the temperature cannot be resolved in such a
simulation. 
Therefore, it is important to measure how large are the thermal
diffusivity in our models of a radiative atmosphere.
The P\'eclet number is a dimensionless number that quantifies the importance
of advective and diffusive term, which is here defined as
\EQ
\Pe=\urms \Hp/\chi,
\EN
where $\Hp$ is a pressure scale height and $\urms$ is rms velocity.
The radiative diffusivity is defined as 
\EQ
\chi=K/\cp\rho,
\label{chi}
\EN
where $K$ is evaluated using \Eq{K-model}.
As we do not solve for a velocity equation in our model, so we use
instead the sound speed, which can be 
related to $\urms$ via the Mach number $\Ma=\urms/\cs$.
The normalized P\'eclet number in our simulation $\Pet$ is then given by
\EQ
\Pet\equiv\Pe/\Ma=\cs\Hp/\chi.
\EN 
As an example, we plot $\Pet$ for Set~C in \Fig{Ra-Pe}.
As we can see in \Fig{Ra-Pe}, $\Pet$ is a large number for
the optically thick part and it decreases as we go toward the
optically thin part.
This can be explained with \Eq{chi}, where $\chi$ is proportional to $K$.
In the optically thin part, $K$ increases, so $\chi$ also increases.
As a result, $\Pet$ decreases.
Furthermore, $\Pet$ is larger for the larger value of $\tkapz$.
 
The quantity $\Pet$ is a measure of the ratio of Kelvin-Helmholtz time
to the sound travel time, $\tau_{\rm sound}=d/\cs$.
In our case, $\tau_{\rm sound}\approx0.1\ks$.
Looking at \Fig{Ra-Pe}, one sees that $\Pet$ is proportional to
$\tkapz$ and thus proportional to the thermal adjustment time.
$\Pet$ depends on $z$, but in the middle of the layer at $z=2\Mm$
we have $\Pet\,\tau_{\rm sound}\approx\tau_{\rm adjust}$.
This time is much longer than the response time to general
three-dimensional disturbances
\citep{Spi87}.

\subsection{The same polytropic index with different $a$ and $b$}

As we can see in \Fig{ab-n}, for a certain value of the polytropic
index, we can choose different combinations of $a$ and $b$.
For each value of $n$ that we have in \Tab{results}, we choose two
different other combinations of $a$ and $b$ with the same value of
$\tkapz=10^5\,\Mm^{-1}\cm^3\g^{-1}$ . 
For example for the polytropic index $n=1$ we choose two other combinations
as $a=0.5$ and $b=1.5$ for one set and $a=2$ and $b=0$ for another
one (see \Tab{test-n}).
We run eight more simulations with the same initial conditions as
in previous runs and we obtain a similar equilibrium
solution for the same polytropic index $n$. 
We calculate the effective temperature and the position where
$\tau\approx1$ as reference parameters with our old runs.
The results are summarized in \Tab{test-n}. For each set of runs with
the same polytropic index, we labeled the runs similarly to those in
\Tab{results}.  
\begin{table}[!t]\caption{Summary of the results for different values of
    $a$ and $b$ with the same polytropic index $n$.
}\vspace{12pt}\centerline{\begin{tabular}{lcccccccc}                            
\hline \hline                                         
Run&$a$ & $b$ & $n$ & $z_{\tau=1}$  & $T_{\rm eff}$\\
\hline
F1&$0.5$ & $-1.9$ & $3.25$ & $5.3$ & 13900\\
F2&  1   & $-3.5$ & $3.25$ & $5.2$ & 13900\\
F3&  2   & $-6.75$& $3.25$ & $5.1$ & 13400\\
\hline
G1&$0.5$ & $0.75$ & $1.5$  & $3.2$ & 16600\\
G2&  1   &    0   & $1.5$  & $2.9$ & 16300\\
G3&  2   & $-1.5$ & $1.5$  & $2.7$ & 16100\\
\hline
H1&$0.5$ & $1.5$  &    1   & $2.6$ & 17100\\
H2&  1   &   1    &    1   & $2.3$ & 17500\\
H3&  2   &   0    &    1   & $2.1$ & 18100\\
\hline
I1&$0.5$ & $4.5$  &  $-1$  & $1.1$ & 21800\\
I2&  1   &   5    &  $-1$  & $0.8$ & 23100\\
I3&  2   &   6    &  $-1$  & $0.6$ & 23700\\
\hline                       
\label{test-n}\end{tabular}}
\tablefoot{
$z_{\tau=1}$ is the position of $\tau\approx1$ in Mm
and $T_{\rm eff}$ is in $\K$.  
For all the runs, $\tkapz=10^5\,\Mm^{-1}\cm^3\g^{-1}$.}
\end{table} 

\begin{figure}[t!]\begin{center}
\includegraphics[width=\columnwidth]{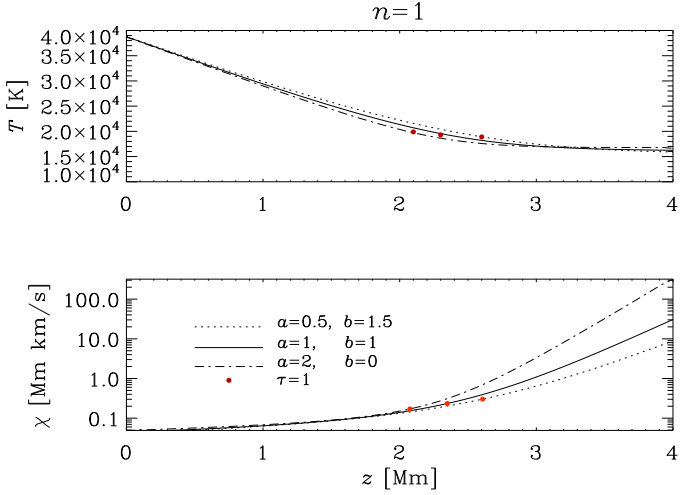}
\end{center}\caption[]{
Profile of $T$ and $\chi$ for Set~H.
In the upper panel, the red dots denote the locations of $\tau=1$.
}\label{Temp-chi}\end{figure}

As we see in \Tab{test-n}, for each set of runs the effective temperature 
does not vary strongly, but there is a systematic behavior.
By increasing the value of $a$, the effective temperature
increases when $n<3/2$ and decreases when $n\geq3/2$, but the surface
is shifted to the lower part of the domain for all sets.
The stratification of temperature and other important properties
of these atmospheres can be explained analogously to those of Sets~A, B, C
and D.
As an example, we plot in \Fig{Temp-chi} the temperature profiles
(upper panel) and
radiative diffusivity $\chi$ (lower panel) for Set~H.  
In the optically thick part, $\chi$ is the same for different
combinations of $a$ and $b$ for the same $n$.
However, in the optically thin part, $\chi$ becomes
larger for larger values of $a$.
This can be explained using \Eqs{K-model}{chi} to show that in the
upper isothermal part, $\chi$ increases with decreasing $\rho$ like
$\rho^{-(a+2)}$.
Thus, we can conclude that, even though $\chi$ is the same and the
solution similar to the optically thick part, there are differences
in the optically thin part.

\subsection{Optically thick case with radiative boundary}
\label{OpticallyThick}

To compare our results with those in the optically thick approximation,
we adopt the radiative boundary condition,
\EQ
-K{\dd T\over\dd z}=\sigmaSB T^4\quad\mbox{on $z=z_{\rm top}$},
\EN
and keep all other conditions the same as in the radiative
transfer calculation, except that $-\nab\cdot\FF_{\rm rad}$
in \Eq{sRT} is replaced by $K\nabla^2T$.
Here, we have assumed $K$ to be constant, so we shall from now on
refer to its value as $K_0$, so our solutions will be
polytropes with constant polytropic index $n=\dd\ln\rho/\dd\ln T$
and constant double-logarithmic temperature gradient
$\nabla=\dd\ln T/\dd\ln p$.

The value of $\nabla=1/(1+n)$ can be computed from the equations
governing hydrothermal equilibrium,
\EQ
{\dd p\over\dd z}=-\rho g,\quad
{\dd T\over\dd z}=-{F_{\rm rad}\over K_0},
\label{dTdz}
\EN
which yields
\EQ
\nabla={\dd\ln T\over\dd\ln p}={p\over T\rho}\,{F_{\rm rad}\over gK_0}
=\nabad\cp\,{F_{\rm rad}\over gK_0}.
\label{nabla_expr}
\EN
Such a model is characterized by choosing values for $n$ and $K_0$.
This is analogous to the case with radiative transfer,
where $n$ and $\tkapz$ are specified, and $\tkapz$ is related to
$K_0$ via \Eq{Kbot}.
Here, it is convenient to define a non-dimensional radiative conductivity as
\EQ
{\cal K}={gK_0\over\cp\sigmaSB T_{\rm bot}^4}\,{\nabla\over\nabad}.
\label{calK}
\EN
The radiative flux is then given by
\EQ
F_{\rm rad}=K_0\,{g\over\cp}\,{\nabla\over\nabad}
={\cal K} \sigmaSB T_{\rm bot}^4,
\label{Frad_calK}
\EN
so we get the temperature at the top immediately as
\EQ
T_{\rm top}=(F_{\rm rad}/\sigmaSB)^{1/4}={\cal K}^{1/4}T_{\rm bot}.
\label{T_top}
\EN
Since the temperatures at top and bottom are now known, the thickness
of the layer cannot be chosen independently and is instead given by
\EQ
d=(T_{\rm bot}-T_{\rm top})\,K_0/F_{\rm rad}
={\cp\,(T_{\rm bot}-T_{\rm top})\over g\,\nabla/\nabad},
\label{dexpression}
\EN
which is equivalent to \Eq{nactual2} used before.
Again, the fact that the value of $d$ cannot be chosen independently
is analogous to the case with radiative transfer, where the thickness
of the optically thick layer with nearly constant $K$ emerges as a result
of the calculation.
In \Tab{tkapz} we present models for the same parameters as in
\Tab{results}, where $\Teff=T_{\rm top}$ in the optically thick model.
In agreement with our radiative transfer calculations,
we have here treated $\tkapz$ (instead of ${\cal K}$) as our
main input parameter (in addition to $n$).
We have used \Eq{Kbot} to convert $\tkapz$ into $K_0$ and then
used \Eq{calK} to compute ${\cal K}$.
It turns out that there is good agreement regarding the values of
$d$, $\rho_{\rm top}$, and $T_{\rm top}$ between the optically thick
approximation using a radiative upper boundary condition and the
radiative transfer calculations.
However, unlike \Fig{Teff-comp}, the data in \Tab{tkapz} show
power law dependence of $T_{\rm eff}$ versus $\tkapz$
with the same exponent of $1/4$ in all cases.
To characterize the strength of density and temperature stratification,
we also list the ratios $\rho_{\rm bot}/\rho_{\rm top}$ and the ratio
of the pressure scale height at the top to the thickness of the layer,
$\xi=\Hp^{\rm top}/d$.
As expected, smaller values of $\xi$ are reached by increasing the
value of $\tkapz$, but even for $\tkapz=10^7\Mm^{-1}\cm^3\g^{-1}$
the smallest values of $\xi$ are 0.03 for $n=3.25$ and 0.08 for $n=1$.
\begin{table}[b!]\caption{
Summary of model parameters as a function of $n$ and $\tilde\kappa_0$
as obtained from the optically thick approximation with radiative upper
boundary condition.
}\vspace{12pt}\centerline{\begin{tabular}{rrrrrrrr}
\hline \hline
$n\;\;$ & $\tilde\kappa_0\;$ & $d\;\;$ & $\rho_{\rm top}\quad\;$ & $T_{\rm top}\;$ & $\xi\;$ &
$\rho_{\rm bot}/\rho_{\rm top}$ \\
\hline
$ 3.25 $&$ 10^{4} $&$ 3.09 $&$ 8.3\;10^{-5} $&$ 24600 $&$ 0.40 $&$   4.5 $\\
$ 3.25 $&$ 10^{5} $&$ 5.40 $&$ 1.3\;10^{-5} $&$ 13800 $&$ 0.13 $&$  28.9 $\\
$ 3.25 $&$ 10^{6} $&$ 6.70 $&$ 2.1\;10^{-6} $&$  7800 $&$ 0.06 $&$ 187~~ $\\
$ 3.25 $&$ 10^{7} $&$ 7.44 $&$ 3.2\;10^{-7} $&$  4400 $&$ 0.03 $&$1220~~ $\\
\hline
$ 1.50 $&$ 10^{4} $&$ 1.37 $&$ 2.2\;10^{-4} $&$ 28100 $&$ 1.04 $&$   1.6 $\\
$ 1.50 $&$ 10^{5} $&$ 2.93 $&$ 8.9\;10^{-5} $&$ 15800 $&$ 0.27 $&$   3.9 $\\
$ 1.50 $&$ 10^{6} $&$ 3.80 $&$ 3.8\;10^{-5} $&$  8900 $&$ 0.12 $&$   9.2 $\\
$ 1.50 $&$ 10^{7} $&$ 4.30 $&$ 1.6\;10^{-5} $&$  5000 $&$ 0.06 $&$  21.8 $\\
\hline
$ 1.00 $&$ 10^{4} $&$ 0.94 $&$ 2.8\;10^{-4} $&$ 29700 $&$ 1.61 $&$   1.3 $\\
$ 1.00 $&$ 10^{5} $&$ 2.25 $&$ 1.4\;10^{-4} $&$ 16700 $&$ 0.38 $&$   2.3 $\\
$ 1.00 $&$ 10^{6} $&$ 2.99 $&$ 8.0\;10^{-5} $&$  9400 $&$ 0.16 $&$   4.1 $\\
$ 1.00 $&$ 10^{7} $&$ 3.41 $&$ 4.5\;10^{-5} $&$  5300 $&$ 0.08 $&$   7.4 $\\
\vspace{-1em}
\label{tkapz}\end{tabular}}
\tablefoot{
The units of dimensional quantities are $[\tkapz]=\Mm^{-1}\cm^3\g^{-1}$,
$[d]=\Mm$, $[\rho_{\rm top}]=\g\cm^{-3}$, and $[T]=\K$.
}\end{table}

\begin{figure*}[t!]
\begin{center}
\includegraphics[width=.9\textwidth]{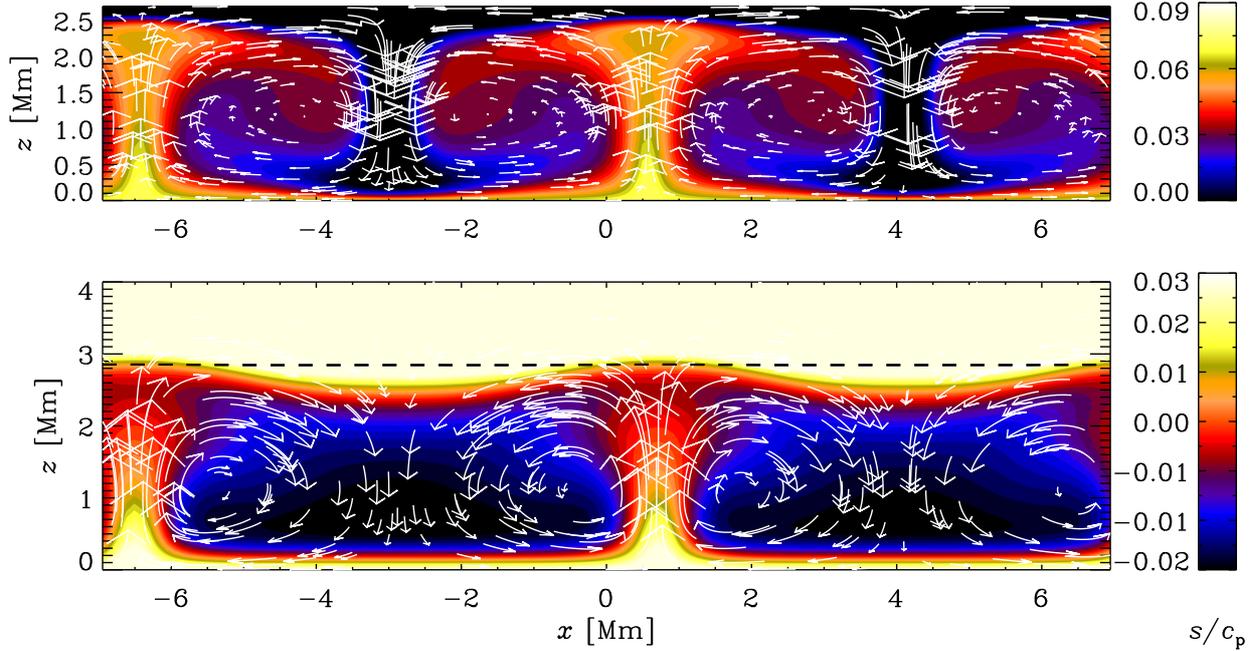}
\end{center}\caption[]{
Comparison of velocity and entropy distribution in two-dimensional
convection using the optically thick approximation with a radiative upper
boundary condition (upper panel) and radiative transfer (lower panel).
In both cases we have $\Pra=100$ and ${\cal K}=0.01$,
corresponding to $\Ra=3.6\;10^4$.
In the lower panel, the dashed line gives the contour $\tau=1$.
In order to compare similar structures in the two plots, we have extended
the color table of the lower panel to slightly more negative values of
$s$ and have clipped it at high values, which are dominated by
the strong increase of $s$ above the $\tau=1$ surface.
}\label{psnap_comp}\end{figure*} 

We emphasize that the only place where the choice of density enters
our calculation is in \Eq{Kbot} when we convert $\tkapz$ into $K_0$.
As already indicated at the end of \Sec{Parameters}, an increase of
$\rho_0$ by some factor is equivalent to an increase of $\tkapz$ by
the same factor.
We note here that $\rho_0$ enters both as the initial density at the
bottom
and in the definition of opacity through \Eq{kappa-ab}.
The latter ensures that the opacity only changes through changes in $\tkapz$,
and not also through changes in $\rho_0$.

\subsection{Convection}
\label{Convection}

We now consider two-dimensional convection and compare again results
from the optically thick approximation using a radiative upper boundary
condition with a calculation using radiative transfer.
The control parameter characterizing onset and amplitude of convection
is the Rayleigh number,
\EQ
\Ra={gd^4\over\nu\chi_{\rm mid}}
\,\left(-{\dd s/\cp\over\dd z}\right)_{\rm mid},
\label{RaDef}
\EN
where $(-\cp^{-1}\,\dd s/\dd z)_{\rm mid}=(\nabla-\nabad)/\Hp^{\rm mid}$
is the superadiabatic gradient of the unstable, non-convecting hydrostatic
reference solution, and $\Hp^{\rm mid}=\nabad\cp T_{\rm mid}/g$ is the
pressure scale height in the middle of the optically thick layer.
Furthermore, we define the Prandtl number as $\Pra=\nu/\chi_{\rm mid}$,
where $\nu$ is assumed to be constant and
$\chi_{\rm mid}$ is the radiative diffusivity in the middle of the
optically thick layer; see \App{AppB} for details and an example.
In \Tab{PrRa} we list the values of the product $\Pra\,\Ra$, as well as $d$
and $\chi_{\rm mid}$ for models with $n=1$ and different values of ${\cal K}$.
We adopt periodic boundary condition in the $x$ direction over
a domain with side length $L_x$.
When we adopt the diffusion approximation we take $z_{\rm top}=d$,
where $d$ is calculated from \Eq{dexpression} and given in \Tab{PrRa}.
The mid-layer is then at $z=d/2$, which is also the case when
using radiative transfer, where the value of $z_{\rm top}$ ($>d$)
was chosen to be sufficiently large
and $a=b=1$ is chosen to yield $n=1$ (see \Tab{results}).

\begin{table}[b!]\caption{
Summary of model parameters as a function of ${\cal K}$ for $n=1$.
}\vspace{12pt}\centerline{\begin{tabular}{cccccc}
\hline \hline
${\cal K}$ & $\Pra\,\Ra$ & $d$ & $\chi_{\rm mid}$ & $K_0$ & $\tkapz$ \\
\hline
$ 5\;10^{-1} $&$ 7.5\;10^{0} $&$ 0.63 $&$ 5.6\;10^{-1} $&$ 6.6\;10^{-6} $&$ 6.8\;10^{3} $\\
$ 2\;10^{-1} $&$\bm{7.1\;10^{2}}$&$ 1.31 $&$ 2.6\;10^{-1} $&$ 2.6\;10^{-6} $&$\bm{1.7\;10^{4}}$\\
$ 1\;10^{-1} $&$ 7.7\;10^{3} $&$ 1.73 $&$ 1.4\;10^{-1} $&$ 1.3\;10^{-6} $&$ 3.4\;10^{4} $\\
$ 5\;10^{-2} $&$ 5.9\;10^{4} $&$ 2.08 $&$ 7.7\;10^{-2} $&$ 6.6\;10^{-7} $&$ 6.8\;10^{4} $\\
$ 2\;10^{-2} $&$\bm{6.6\;10^{5}}$&$ 2.46 $&$ 3.3\;10^{-2} $&$ 2.6\;10^{-7} $&$ 1.7\;10^{5} $\\
$\bm{1\;10^{-2}}$&$\bm{3.6\;10^{6}}$&$\!\bm{2.70}\!$&$\!\bm{1.8\;10^{-2}}\!$&$\bm{1.3\;10^{-7}}$&$\bm{3.4\;10^{5}}$\\
$ 5\;10^{-3} $&$ 1.9\;10^{7} $&$ 2.89 $&$ 9.1\;10^{-3} $&$ 6.6\;10^{-8} $&$ 6.8\;10^{5} $\\
$ 2\;10^{-3} $&$ 1.5\;10^{8} $&$ 3.11 $&$ 3.8\;10^{-3} $&$ 2.6\;10^{-8} $&$ 1.7\;10^{6} $\\
$ 1\;10^{-3} $&$ 6.9\;10^{8} $&$ 3.24 $&$ 1.9\;10^{-3} $&$ 1.3\;10^{-8} $&$ 3.4\;10^{6} $\\
$ 5\;10^{-4} $&$ 3.1\;10^{9} $&$ 3.35 $&$ 9.9\;10^{-4} $&$ 6.6\;10^{-9} $&$ 6.8\;10^{6} $\\
$ 2\;10^{-4} $&$ 2.2\;10^{10} $&$ 3.47 $&$ 4.1\;10^{-4} $&$ 2.6\;10^{-9} $&$ 1.7\;10^{7} $\\
$ 1\;10^{-4} $&$ 9.5\;10^{10} $&$ 3.55 $&$ 2.1\;10^{-4} $&$ 1.3\;10^{-9} $&$ 3.4\;10^{7} $\\
\label{PrRa}\end{tabular}}
\tablefoot{
The values discussed and used in this paper are shown in bold face
and are valid for any value of $\Pra$.
The units are $[d]=\Mm$, $[\chi_{\rm mid}]=\Mm\km\s^{-1}$,
$[K_0]=\g\cm^{-3}\km^3\s^{-3}\Mm\K^{-1}$, and
$[\tkapz]=\Mm^{-1}\cm^3\g^{-1}$.
}\end{table}

We determine the critical value for the onset of convection by calculating
the rms velocity in the domain, $\urms$, for different values of ${\cal K}$
and extrapolate to $\urms\to0$.
For the final to initial bottom density ratio, $\rho_{\rm bot}/\rho_0$,
we use \Eq{DensityRatio} derived in \App{BottomDensityRatio}.
Since ${\cal K}$ is proportional to $\chi$ and $\chi_{\rm mid}$, we can
compute the product $\Pra\,\Ra$ using \Eq{RaDef}.
It turns out that for $\Pra=1$, the critical value is at
${\cal K}\ga 0.2$ (corresponding to $\Ra\la710$) in the optically thick
approximation.
This corresponds to $\tkapz=1.7\;10^4\Mm^{-1}\cm^3\g^{-1}$,
which is too small to obtain a proper polytropic lower part.
Therefore we choose in the following $\Pra=100$, in which cases the
critical value for the onset of convection is at ${\cal K}\ga 0.02$
(corresponding to $\Ra\la6600$).

\begin{figure}[t!]\begin{center}
\includegraphics[width=\columnwidth]{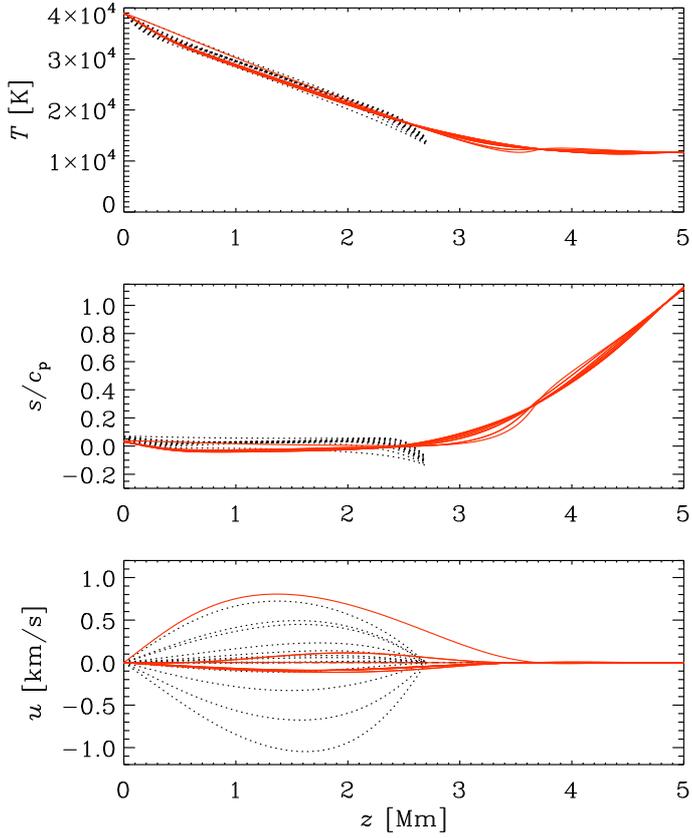}
\end{center}\caption[]{
Comparison of vertical temperature, entropy, and velocity profiles at
different $x$ positions for the model shown in \Fig{psnap_comp}.
The black dotted lines refer to the run with 
diffusion approximation and radiative upper boundary
condition and the red solid lines to the run with radiative transfer.
}\label{pTU_profiles_comp}\end{figure}

In the following, we take ${\cal K}=0.01$, so $\Ra=3.6\;10^4$, $d=2.70\Mm$,
$\nu=1.8\Mm\km\s^{-1}$ (corresponding to $\nu=1.8\;10^{13}\cm^2\s^{-1}$),
and $\tkapz=3.4\;10^5\Mm^{-1}\cm^3\g^{-1}$; see the sixth row of \Tab{PrRa}.
We choose $L_x=14\Mm$, which is large enough to accommodate two convection
cells into the domain; see \Fig{psnap_comp}.
The $\tau=1$ surface in the radiative transfer calculation
agrees approximately with the height expected from the optically thick
models using a radiative upper boundary condition.
The flow is only weakly supercritical and therefore not very vigorous,
which is also reflected by the fact that the $\tau=1$ surface is nearly flat.
In the radiative transfer calculation, the specific entropy increases sharply
with height above the $\tau=1$ surface.
We note also that the characteristic narrow downdrafts of the optically
thick calculation are now much broader when radiation transfer is used.
Furthermore, the expected entropy minimum near the surface is
virtually absent in the latter case; see also the middle panel
of \Fig{pTU_profiles_comp}.
This is because near $z=d$, the local value of $\chi$ is rather large
(see the lower panel of \Fig{Temp-chi}), so the thickness of the
thermal boundary layer becomes comparable to $d$ itself.

\begin{figure}[t!]
\begin{center}
\includegraphics[width=\columnwidth]{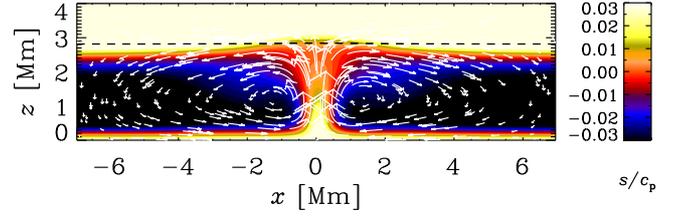}
\end{center}\caption[]{
Similar to the lower panel of \Fig{psnap_comp}, but at a later time
($t=500\ks$), when the solution has switched into a single-cell
configuration.
}\label{psnap_single_Pr100}\end{figure}

In the model using the diffusion approximation, the temperature
variations are much larger than in the model with radiative transfer;
see \Fig{pTU_profiles_comp}.
In the latter case, the velocities overshoot into the upper stably stratified
layer, and the downflows are much broader and therefore slower than
in the diffusion approximation.

It turns out that the solution with radiative transfer
shown in \Fig{psnap_comp} is quasi-stable until about $350\ks$
($\approx4$\,days) and then switches into a single-cell configuration
with a fairly isolated updraft; see \Fig{psnap_single_Pr100}.
We have seen similar behavior in other cases with radiation too, and
it is possible that this is a consequence of our setup.
Firstly, the restriction to two-dimensional convection is a serious artifact.
Secondly, the assumption of a fixed temperature at the bottom was a mathematical
convenience, but it is not physical motivated.

\section{Increasing the density contrast}

As alluded to in the introduction, the inclusion of the physics of
radiation implies the occurrence of $\sigmaSB$ as an additional physical
constant that couples the resulting temperature and density contrasts
to changes in the Rayleigh number.
Given that we have already made other simplifications such as the
negligence of hydrogen ionization, we end up with rather small
density contrasts of less than ten when $n=1$, as seen in \Tab{tkapz}.
By considering $\sigmaSB$ an adjustable parameter, we can
alleviate this constraint.
We demonstrate this in \Tab{contrast}, where we increase
the value of $\sigmaSB$ from its physical value of
$5.67\;10^{-20}\g\cm^{-3}\km^3\s^{-3}\K^{-4}$ by eight orders
of magnitude, keeping however $K_0$ fixed.
We have chosen here
$K_0=1.3\;10^{-7}=\g\cm^{-3}\km^3\s^{-3}\Mm\K^{-1}$,
which corresponds to the model in the sixth row of \Tab{PrRa}.
Since $\sigmaSB$ enters the definition of ${\cal K}$,
its value is now no longer the same, even though $K_0$ is.
The Rayleigh numbers change slightly, because they
depend on the values of density and temperature
in the middle of the domain, which do of course change.

\begin{table}[b!]\caption{
Density contrast and other model parameters as a function of
$\sigmaSB$ for $n=1$ and $K_0=1.3\times10^{-7}$.
}\vspace{12pt}\centerline{\begin{tabular}{rlrrrc}
\hline \hline
$\sigmaSB\quad\;\;$ & $\;\Pra\,\Ra$ & $d\;\;\;$ & $\chi_{\rm
        mid}\quad$ & $\rho_{\rm bot}/\rho_{\rm top}\!\!\!\!$ & $\tkapz$ \\
\hline
$ 5.67\;10^{-20} $&$ 3.63\;10^{6} $&$ 2.70 $&$ 1.8\;10^{-2} $&$   3.2 $&$ 3.4\;10^{5} $\\
$ 5.67\;10^{-18} $&$ 9.45\;10^{6} $&$ 3.55 $&$ 2.1\;10^{-2} $&$  10.0 $&$ 3.4\;10^{7} $\\
$ 5.67\;10^{-16} $&$ 1.23\;10^{7} $&$ 3.82 $&$ 2.2\;10^{-2} $&$  31.6 $&$ 3.4\;10^{9} $\\
$ 5.67\;10^{-14} $&$ 1.33\;10^{7} $&$ 3.90 $&$ 2.2\;10^{-2} $&$ 100.0 $&$ 3.4\;10^{11}$\\
$ 5.67\;10^{-12} $&$ 1.36\;10^{7} $&$ 3.93 $&$ 2.2\;10^{-2} $&$ 316.2 $&$ 3.4\;10^{13}$
\label{contrast}\end{tabular}}
\tablefoot{
The units are $[\sigmaSB]=\g\cm^{-3}\km^3\s^{-3}\K^{-4}$,
$[d]=\Mm$, $[\chi_{\rm mid}]=\Mm\km\s^{-1}$, and
$[\tkapz]=\Mm^{-1}\cm^3\g^{-1}$.
}\end{table}

Large density contrasts are one of the important ingredients
in modeling the physics of sunspot formation by surface effects
such as the negative effective magnetic pressure instability;
see \cite{BKR13} for a recent model.
Including radiation into such still rather idealized models and
to study the relation to other competing or corroborating
mechanisms such as the one of \cite{KM00} was
indeed an important motivation behind the work of the present paper.

\section{Conclusions}

The inclusion of radiative transfer in a hydrodynamic code
provides a natural and physically motivated way of placing
an upper stably stratified layer on top of an optically thick layer
that may be stably or unstably stratified,
which of the two depends on the opacity.
Using a Kramers-like opacity law with freely adjustable exponents
on density and temperature yields polytropic solutions for certain
combinations of the exponents $a$ and $b$.
The prefactor in the opacity law determines essentially the values
of the P\'eclet and Rayleigh numbers.
However, in contrast to earlier studies of convection in polytropic
layers, the temperature contrast is no longer a free parameter and
increases with increasing Rayleigh number---unless one considers
the Stefan--Boltzmann `constant' as an adjustable parameter.
The physical values of the prefactor on the opacity are much larger than those
used here, but larger prefactors lead to values of the radiative
diffusivity that become eventually so small that temperature fluctuations
on the mesh scale cannot be dissipated by radiative diffusion.
In previous work \citep{Nor82,SLK89,VSSCE05,HNSS07,Freytag12},
this problem has been avoided by applying numerical diffusion or
using numerical schemes that dissipate the energy when and where needed.
However, this may also suppress the possibility of physical instabilities
that we are ultimately interested in.
This motivates the investigation of models with prefactors in the
Kramers opacity law that are manageable without the use of numerical
procedures to dissipate energy artificially.

It turns out that in all cases with $a$ and $b$ such that $n>-1$,
the stratification corresponds to a polytrope with index $n$
below the photosphere and to an isothermal one above it.
This was actually expected given that such a solution has previously
been obtained analytically in the special case of constant $\kappa$
(corresponding to $a=b=0$); see \cite{Spi06}.
On the other hand, the isothermal part was apparently not present
in the simulations of \cite{Edw90}.

Contrary to the usual polytropic setups \citep[e.g.,][]{Bra96},
the temperature contrast is now no longer an independent parameter
but it is tied essentially to the Rayleigh number.
Significant temperature contrasts can only be achieved
at large Rayleigh numbers, which corresponds to a large prefactor
in front of the opacity.
While this aspect can be reproduced already with polytropic models
using the radiative boundary conditions, there are some surprising
differences between the two.
Most important is perhaps the fact that the specific entropy must
increase above the unstable layer, while with a radiative boundary
condition the specific entropy always decreases.
Although the differences in the resulting temperature profiles are
small, there are major differences in the flows speeds in the two cases.
We also find that at late times the convection cells in simulations
with full radiative transport tend to merge into larger ones.
Whether or not this is an artefact of our restriction to two-dimensional
flows remains open.
In this connection, we should also point out the presence of a geometric
correction factor in front of the radiative heating and cooling term
in \Eq{Sum} that is needed to reproduce the correct cooling rate,
but it does not affect the steady state solution.

Comparing with realistic simulations of the Sun, there is not really
an isothermal part, but a pronounced sudden drop in temperature followed
by a continued decrease in temperature \citep[see, e.g.,][]{SN98}.
On the other hand, in our simulations there is no jump in the
temperature profile
near the surface and the atmosphere changes smoothly from polytropic to
isothermal.
We suspect that the reason for this difference is that in our models
ionization effects are ignored, while in the solar atmosphere
the degree of ionization of hydrogen increases with depth.
In the Sun, the density decreases significantly from the upper part
of the convection zone as we go to the photosphere.
This makes the opacity smaller and the atmosphere in the photosphere
becomes transparent.
At the height where the ionization temperature of hydrogen is reached,
the $H^-$ 
opacity becomes important, which is not included in our simulations.   
The radiative heat conductivity in our simulations is found to be
constant
throughout the optically thick part and then increases sharply in the
optically thin part.
Solving this in the optically thick approximation, which has sometimes
been done, becomes computationally expensive and even unphysical, so
radiative transfer becomes a viable alternative for studying layers
that are otherwise polytropic in the lower part of the domain.

\begin{acknowledgements}

We are indebted to Tobi Heinemann for his engagement in implementing
the radiative transfer module into the {\sc Pencil Code}.
We also thank the referee for many helpful comments that have led
to improvements of the manuscript.
This work was supported in part by the European Research Council under the
AstroDyn Research Project No.\ 227952, and
by the Swedish Research Council under the project grants
621-2011-5076 and 2012-5797.
We acknowledge the allocation of computing resources provided by the
Swedish National Allocations Committee at the Center for
Parallel Computers at the Royal Institute of Technology in
Stockholm and the National Supercomputer Centers in Link\"oping, the High
Performance Computing Center North in Ume\aa,
and the Nordic High Performance
Computing Center in Reykjavik.
\end{acknowledgements}

\appendix
\section{Cooling rate and correction factor}
\label{AppA}

The purpose of this appendix is to show that for
one-dimensional temperature perturbations, the correct
cooling rates are obtained with just two rays
if the $D/3$ correction factor in \Eq{Sum} is applied.
Similar considerations apply also to the case of
two-dimensional problems.
Cooling rates are important for understanding temporal aspects
such as the approach to the final state (\Sec{Approach}) or the thermal
adjustment time (\Sec{t-adjust}).
Thus, the equilibrium solutions discussed in the other sections
are not affected by the following considerations.

The source of the problem lies in the fact that the $4\pi$ angular
integration in \Eq{fff} becomes inaccurate in one dimension and
dependent on the optical thickness.
In the optically thick regime, the diffusion approximation holds,
so the cooling rate is proportional to $K$, which has a $1/3$ factor
in \Eq{K-model}.
In one dimension, one uses only the two rays in the vertical direction,
so one misses the $1/3$ factor and has to apply it afterwards to
account for the ``redundant'' rays in the other two coordinate directions
that show no variation.
This is what is done in \Eq{Sum}.
However, in the optically thin limit, the mean free path becomes infinite
and cooling is now possible in all three directions.
In that case, the one-dimensional approximation is not useful.
To explain this in more detail, we begin by considering first the general case
of three-dimensional perturbations with wavevector $\kk$ \citep{Spi57}.
In that case, one can use the Eddington approximation to solve the transfer equation
for the mean intensity, $J=\int I\,\dd\Omega/4\pi$,
\begin{equation}
\onethird\,(\ell\nabla)^2J=J-S,
\label{Eddington}
\end{equation}
so the cooling rate
(for three-dimensional perturbations) is \citep{US66,Edw90}
\begin{equation}
\lambda_{\rm3D}={16\sigmaSB T^3\over\rho\cp}\,
{\kappa\rho\kk^2\over3\kappa^2\rho^2+\kk^2}.
\label{decay3D}
\end{equation}
It is convenient to introduce here a photon diffusion speed as
\begin{equation}
c_\gamma=16\sigmaSB T^3/\rho\cp
\end{equation}
and to write \Eq{decay3D} in the form
\begin{equation}
\lambda_{\rm3D}={c_\gamma\ell\kk^2/3\over1+\ell^2\kk^2/3},
\label{decay3Db}
\end{equation}
where $c_\gamma\ell/3=\chi$ is the radiative diffusivity,
as defined in \Eq{chi}, and $\ell=1/\kappa\rho$ is the
local mean-free path of photons.

Solving \Eq{RT-eq} for two rays corresponds to solving
\Eq{Eddington} without the $1/3$ factor.
We would then obtain \Eq{decay3Db} without the two $1/3$ factors.
This would evidently violate the well-known cooling rate $\chi\kk^2$ in
the optically thick limit, but in the optically thin limit it would be in
agreement with \Eq{decay3Db}, because the two $1/3$ factors would cancel
for large values of $\ell$.
However, we have to remember that temperature perturbations are
here assumed one-dimensional, so the intensity can only vary in the
$z$ direction, while the rays still go in all three directions.
This means that under the sum in \Eq{Sum} only one third of the
$I-S$ terms give a contribution, and that the cooling rate
is therefore
\begin{equation}
\lambda_{\rm1D}={c_\gamma\ell k_z^2/3\over1+\ell^2 k_z^2},
\label{decay1D}
\end{equation}
which has now only a single $1/3$ factor.
Likewise, if we had two-dimensional perturbations such as in
two-dimensional convection considered in \Sec{Convection},
only $2/3$ of the terms under the sum in \Eq{Sum} would contribute.
However, in a two-dimensional radiative transfer calculation,
the additional $1/3$ would be absent, which explains the
$D/3$ correction factor with $D=2$ in this case.

\begin{figure}[t!]\begin{center}
\includegraphics[width=\columnwidth]{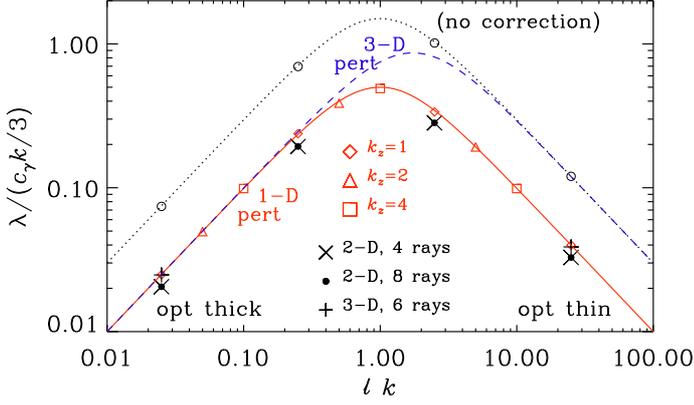}
\end{center}\caption[]{
Dependence of the cooling rates computed from models with different
values of $\tkapz$ (from $10^2$ to $10^5$
$\Mm^{-1}\cm^3\g^{-1}$) and $k_z$ (=1, 2, and 4 
indicated by diamonds, triangles, and squares, respectively).
2-D models with four and eight rays are indicated by crosses and circles,
respectively, while 3-D models with six rays are shown as plus signs.
The red solid line corresponds to \Eq{decay1D},
the dashed blue line to \Eq{decay3Db}, and the dotted line
with open circles to the case without correction factor.
}\label{pdecay_comp}\end{figure}

We have verified that with the correction factor in place,
the code now yields the same cooling rates in both the optically
thick and thin regimes, regardless of the numbers of rays used.
This is shown in \Fig{pdecay_comp}, where we plot cooling
rates for different values of $\tkapz$ in a domain of size
$2\pi$ (in Mm), so the smallest wavenumber is $1\Mm^{-1}$.
With $\rho=4\times10^{-4}\g\cm^{-3}$ the photon mean-free path varies
from 0.025 to 25 $\Mm$ as $\tkapz$ is decreased from $10^5$ to
$10^2$ $\Mm^{-1}\cm^3\g^{-1}$.
For the Kramers opacity, we use the exponents $a=1$ and $b=0$.
(No gravity is included here, so there would be no convection.)
The temperature is $38,968\K$, as before, which yields
$c_\gamma=3.87\km\s^{-1}$ for the photon diffusion speed.
There is excellent agreement between 1-D cases with correction
factor and the 3-D calculation (with one-dimensional perturbation).
However, the 2/3 correction factor in the 2-D calculation
(both with four and with eight rays) seems to be systematically off
and should instead by around 0.8 for better agreement.
However, as discussed before, the correction factor does not
affect the steady state and therefore also not the results
presented in \Sec{Convection}.
The diffusion approximation would imply
$\lambda=(c_\gamma k/3)\ell k=\chi k^2$,
which corresponds to the diagonal in \Fig{pdecay_comp}
and agrees with the red solid line for $\ell k\la0.5$.

For three-dimensional perturbations, the correct cooling rate
in the optically thin regime is three times faster than for
one-dimensional perturbations.
This is because now the radiation goes in all three directions.
Solutions to three-dimensional perturbations clearly cannot be reproduced
in less than three dimensions.
However, for one-dimensional perturbations, the correct cooling rate
is now obtained with a one-dimensional calculation {\em both} in the
optically thin and thick regimes.

\section{Expressions for $\Pra\,\Ra$ and $\chi_{\rm mid}$}
\label{AppB}

In \Tab{PrRa} we listed the values of $\Pra\,\Ra$ and $\chi_{\rm mid}$
in the middle of the layer.
The purpose of this appendix is to give the explicit expressions
and to demonstrate the calculation with the help of an example.
Since $n=1$ was assumed, we have $\nabla=(1+n)^{-1}=1/2$.
Considering the case ${\cal K}=0.01$, \Eqss{Frad_calK}{dexpression} yield
$F_{\rm rad}=0.00131\g\cm^{-3}\km^3\s^{-3}$, $T_{\rm top}=12320\K$,
and $d=2.70\Mm$.
Next, given that the temperature varies linearly, we compute the
mid-layer temperature as $T_{\rm mid}=\half(T_{\rm top}+T_{\rm bot})=25600\K$.
This allows us to compute
$\rho_{\rm mid}=\rho_{\rm bot}\,(T_{\rm mid}/T_{\rm bot})^n
=2.2\;10^{-4}\g\cm^{-3}$, where $\rho_{\rm bot}=3.3\;10^{-4}\g\cm^{-3}$
is smaller than $\rho_0$ by a factor $\rho_{\rm bot}/\rho_0=0.83$;
see \App{BottomDensityRatio}.
Thus,
$\chi_{\rm mid}=F_{\rm rad}/(\rho_{\rm mid}g\nabla/\nabad)=0.0175\Mm\km\s^{-1}$,
as well as $\Hp^{\rm mid}=\nabad\cp T_{\rm mid}/g=1.30\Mm$.
This yields $\Pra\,\Ra=gd^4/\chi_{\rm mid}^2(\nabla-\nabla_{\rm ad})/
H_p^{\rm mid}=3.6\;10^6$, where $\nabla-\nabla_{\rm ad}=0.1$.

\section{Final to initial bottom density ratio}
\label{BottomDensityRatio}

Initially, the stratification is isothermal, so the density is given by
$\rho(z)=\rho_0\exp(-z/\Hp^{\rm bot})$ and the initial surface density is
\EQ
\Sigma_{\rm ini}=\int_0^d\rho(z)\,\dd z=\rho_0 \Hp^{\rm bot}
\left[1-\exp(-d/\Hp^{\rm bot})\right].
\label{Sigma_ini}
\EN
In the final state, the stratification is polytropic, so the density is given
by $\rho(z)=\rho_{\rm bot}[T(z)/T_{\rm bot}]^n$ and the surface density is
\EQ
\Sigma_{\rm fin}=\rho_{\rm bot}\int_0^d
\left[{T(z)\over T_{\rm bot}}\right]^n{\dd z\over\dd T}\,\dd T.
\label{Sigma_fin3}
\EN
Here, $\rho_{\rm bot}$ is the bottom density of the final state,
which is different from the initial value $\rho_0$,
as explained in \Sec{SimulationStrategy}.
Integrating \Eq{Sigma_fin3} and using $\dd z/\dd T=K_0/F_{\rm rad}$
from \Eq{dTdz} yields
\EQ
\Sigma_{\rm fin}={\rho_{\rm bot}\over n+1}
\left[1-\left({T_{\rm top}\over T_{\rm bot}}\right)^{n+1}\right]
{K_0 T_{\rm bot}\over F_{\rm rad}}.
\EN
Using \Eq{nabla_expr} together with $\nabla=1/(1+n)$ and
$\Hp^{\rm bot}=\nabad\cp T_{\rm bot}/g$, we have
\EQ
\Sigma_{\rm fin}=\rho_{\rm bot}\Hp^{\rm bot}
\left[1-\left({T_{\rm top}\over T_{\rm bot}}\right)^{n+1}\right].
\label{Sigma_fin}
\EN
Using mass conservation, we have $\Sigma_{\rm fin}=\Sigma_{\rm ini}$,
so we obtain from \Eqs{Sigma_ini}{Sigma_fin}
\EQ
{\rho_{\rm bot}\over\rho_0}={1-e^{-d/\Hp^{\rm bot}}\over
1-(T_{\rm top}/T_{\rm bot})^{n+1}}.
\label{DensityRatio}
\EN
for the final to initial bottom density ratio.


\end{document}